\documentclass[ reprint,
bibnotes,
 amsmath,amssymb,
 aps,
pra,
]{revtex4-2}
\usepackage{lipsum}
\usepackage{graphicx}%
\usepackage{soul} %
\usepackage{dcolumn}%
\usepackage{bm}%
\usepackage{cancel}
\usepackage{comment}
\usepackage{hyperref}%
\usepackage{physics} %
\usepackage[usenames, dvipsnames]{color}
\hypersetup{
    colorlinks=true,
    linkcolor=Blue,
    citecolor=Blue,
    filecolor=Blue,      
    urlcolor=Blue,
    }

\begin{document}

\title{Landau Levels on the Surface of a Cube}%
\author{Almila Kura}

 \email{almila.kura@bilkent.edu.tr}

\author{Bayram  Tekin}
 \email{bayram.tekin@bilkent.edu.tr}
\author{M. \"O. Oktel}
 \email{oktel@fen.bilkent.edu.tr}
\affiliation{Department of Physics, Bilkent University, Ankara, 06800, Türkiye}

\date{\today}%

\begin{abstract}
We study the quantum mechanics of a charged particle confined to the surface of a cube enclosing a magnetic monopole. The magnetic field is chosen to have a constant magnitude on each face and to point along the outward normal, preserving the rotational symmetry of the cube. We formulate the continuum problem using two gauge patches on the cube surface and show that consistency of the wavefunction gives the Dirac quantization condition. Since an explicit vector potential does not remain invariant under ordinary rotations, we construct gauge-modified rotation operators and use them to classify the eigenstates. Even monopole charges are described by the irreducible representations of the cubic rotation group $O$, while odd monopole charges require the spinorial representations of the binary octahedral group $2O$. We compute the spectrum with a gauge-covariant finite-difference discretization and find Landau-level-like manifolds whose degeneracies are split by the discrete cubic symmetry. We also study the corresponding tight-binding Hofstadter problem on the discretized cube. The resulting spectrum contains the usual magnetic subband structure together with additional gap states localized near the cube corners.

\end{abstract}

\maketitle

\section{Introduction}

Quantum mechanics on compact surfaces is shaped not only by local dynamics but also by global geometry. Even when a particle moves freely on each small patch of the surface, the way these patches are joined determines the allowed energies, degeneracies, and wavefunctions. The spectrum therefore reflects the geometry and symmetry of the underlying space.

Polyhedral surfaces provide a simple setting in which this connection can be studied. Each face is flat, so the particle is locally described by the usual free-particle Schrödinger equation. The nontrivial part of the problem lies in how the faces are connected at edges and vertices. The regular tetrahedron \cite{belin_quantum_2019} provides an exactly solvable example, and the quantum particle on the surface of a cube has recently been analyzed \cite{cidlinsky_quantum_2024} by classifying the eigenstates according to the irreducible representations of the cubic symmetry group.

A natural extension is to add a magnetic field while preserving the  cube's symmetry and homogeneity on each face. The appropriate field has a constant magnitude on each face and is directed along the outward normal. Such a field carries the net magnetic flux of a monopole enclosed by the cube. The resulting problem remains locally simple on each face, but the magnetic field introduces a nontrivial gauge structure. Quantum mechanics is then determined by the combined effects of the cube geometry, the magnetic flux, and the cubic symmetry. 

Closely related monopole problems have previously been studied on polyhedral graphs. Tight-binding models on the vertices of regular polyhedra exhibit flux-dependent spectra and monopole-induced degeneracies \cite{avishai_tight-binding_2008,oktel_spectrum_2012}, while discrete magnetic Laplacians on regular polyhedra have been formulated using the representation theory of binary polyhedral groups \cite{kemp_discrete_2013}. The present work differs from these graph-based constructions by treating the particle as moving continuously over the faces of the cube. The same face discretization is later used to define a distinct tight-binding problem.

This problem is also connected to several familiar themes in quantum physics. The monopole flux through a closed surface leads to the Dirac quantization condition \cite{dirac_quantised_1931}, while the Wu--Yang construction gives a nonsingular two-patch formulation and the corresponding monopole harmonics on the sphere \cite{wu_dirac_1976}. On the sphere, each Landau-level manifold has finite degeneracy, providing the basis for Haldane's spherical geometry used in studies of quantum Hall systems \cite{haldane_fractional_1983}. More broadly, synthetic gauge fields \cite{goldman_light-induced_2014} and advances in light-shaped optical confinement of ultracold gases \cite{navon_quantum_2021} make geometrically structured magnetic problems relevant to quantum simulation.

In this work, we study the quantum mechanics of a charged particle on the surface of a cube enclosing such a monopole. We first formulate the continuum problem using two gauge patches on the cube surface. This gives a direct analogue of the Wu--Yang construction \cite{wu_dirac_1976}, adapted to a polyhedral surface. The consistency condition between the two patches gives the allowed monopole strengths and fixes the flux quantization condition in this geometry.

We then construct the symmetry operations of the magnetic problem. Although the magnetic field has the full rotational symmetry of the cube, a particular vector potential does not. Each rotation must therefore be accompanied by a gauge transformation. These gauge-modified rotations commute with the Hamiltonian and provide the correct symmetry classification of the eigenstates. For even monopole charge, the states are classified by the ordinary cubic rotation group $O$, while for odd monopole charge, the relevant group is its double cover $2O$.

The spectrum is obtained numerically using a gauge-covariant discretization of the Hamiltonian on the cube surface. This method reproduces the known zero-field spectrum \cite{cidlinsky_quantum_2024} and allows us to track the low-energy states as the monopole strength increases. The resulting levels form Landau-level-like manifolds, but their degeneracies are split by the reduced symmetry of the cube. In particular, the sequence of low-energy degeneracies is governed by the irreducible representations of $O$ or $2O$, rather than by the full rotational symmetry of the sphere. We also construct representative wavefunctions adapted to the cubic symmetry.

Finally, we use the same discretized cube to study the corresponding tight-binding Hofstadter problem \cite{hofstadter_energy_1976}. The spectrum shows the expected magnetic subband structure, but also contains additional states in the gaps when compared with the usual torus geometry. These states are localized near the cube vertices and arise from the special coordination of the corner sites. This comparison separates the local square-lattice Hofstadter physics from the additional effects caused by the polyhedral geometry.

The paper is organized as follows. In Sec.~II we define the continuum Hamiltonian, the monopole field on the cube, and the two gauge patches used to describe it. In Sec.~III we construct the gauge-modified symmetry operations and explain the classification in terms of $O$ and $2O$. Section~IV describes the gauge-covariant numerical discretization and the implementation of the symmetry operators on the lattice. Section~V presents the continuum spectra, symmetry labels, and representative wavefunctions. In Sec.~VI we discuss the tight-binding Hofstadter problem on the cube and the origin of the corner-localized gap states. Section~VII contains our conclusions.

\section{The Model}
\label{sec:model}

We consider a particle of mass $m$ and electric charge $q$, constrained to move on the surface of a cube of side $L$. We adopt an intrinsic formulation in which the cube surface is treated as a two-dimensional manifold assembled from six flat faces. This should be distinguished from thin-layer confinement near an embedded surface, which generally produces an additional curvature-dependent geometric potential \cite{da_costa_quantum_1981}. The Hamiltonian is
\begin{equation}
\tilde{\cal H}
=
-\frac{\hbar^2}{2m}
\left(
\tilde{\vec \nabla}_S
-i\frac{q}{\hbar}\tilde{\vec A}(\tilde{\vec r})
\right)^2,
\end{equation}
where $\tilde{\vec\nabla}_S$ denotes the tangential gradient on the surface ($S$ denoting the surface and tildes will soon be dropped out after rescaling) and $\tilde{\vec A}$ is the tangential vector potential. The full vector potential defines the magnetic field
$\tilde{\vec B}=\tilde{\vec \nabla}\times\tilde{\vec A}$, with
$ \tilde{\vec \nabla}= \hat{x} \frac{\partial}{\partial{\tilde{x}}} +\hat{y}\frac{\partial}{\partial{\tilde{y}}} +\hat{z}\frac{\partial}{\partial{\tilde{z}}} $. 

We scale all lengths by the cube side, $\vec{r}=\tilde{\vec r}/L$, giving $\vec{\nabla}=L \tilde{\vec \nabla}$ and arrive at the scaled Hamiltonian
\begin{equation}
{\cal H}=\tilde{\cal{H}}/\left( \frac{\hbar^2}{2 m L^2}\right) = - \left( \vec{\nabla} - i \frac{qL}{\hbar}\tilde{\vec{A}}\right)^2.
\end{equation}
We will report all energies in terms of the scaled Hamiltonian, i.e., in units of $\left( \frac{\hbar^2}{2 m L^2}\right)$. Similarly, the scaled vector potential is $\vec{A}=\frac{q L}{\hbar} \tilde{\vec{A}}$. In terms of the scaled variables, the Hamiltonian is simply
\begin{equation}
\label{eqn:dimensionless hamiltonian}
H=-\left(\vec{\nabla}-i \vec{A}(\vec{r})\right)^2.
\end{equation}

The particle is confined to the surface of the cube. On each face, the surface is flat, and the Hamiltonian is locally the usual two-dimensional kinetic-energy operator with minimal coupling. The nontrivial part of the problem is the way the six faces are joined at the edges. The wavefunction must be continuous across every edge, and the derivatives must be matched after identifying the two adjacent faces. These matching conditions define the quantum problem on the cube surface.

\begin{figure}[hbt!]
    \centering
    \includegraphics[width=0.9\linewidth]{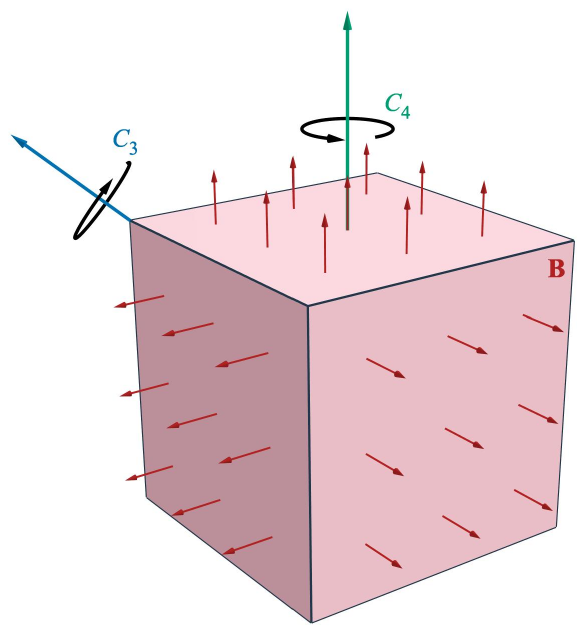}
    \caption{Schematic of a cube enclosing a magnetic monopole. The red arrows show the constant outward face-normal magnetic field. The fourfold axis $C_4$, passing through the centers of opposite faces, and the threefold axis $C_3$, lying along a body diagonal, generate the proper octahedral rotation group $O$. Other rotations, such as $C_2$, are obtained as products of these generators.}
    \label{fig:cube and generators}
\end{figure}

We choose a magnetic field that has a constant magnitude on each face and is directed along the outward normal,as schematically displayed in Fig.~(\ref{fig:cube and generators}). This field treats all faces equally and preserves the cube's rotational symmetries. Since the total outward flux through the closed surface is nonzero, this field represents a magnetic monopole enclosed by the cube. It is not the radial field of a point Dirac monopole. Rather, it is a ``fat monopole'' field, chosen so that the local field strength is the same on every point of the surface while the full cubic symmetry is preserved.

The magnetic flux through an oriented surface patch $S$, whose boundary is the closed contour $C=\partial S$, is 
\begin{equation}
\Phi_B(S)
=
\int_S \tilde{\vec B}\cdot d\tilde{\vec S}
=
\oint_{C}\tilde{\vec A}\cdot d\tilde{\vec\ell}
=
\frac{\hbar}{q}
\oint_{C}\vec A\cdot d\vec\ell.
\end{equation}
Consequently,
\begin{equation}
\label{eqn:flux quantum}
\oint_{C}\vec A\cdot d\vec\ell
=
2\pi\frac{\Phi_B(S)}{\phi_0},
\end{equation}
where $\phi_0=h/q$ is the flux quantum. The total scaled flux outgoing from a face is $B$, as all lengths are scaled by the side of the cube. The total flux outgoing from the cube is, thus, $6 B$.

\begin{figure}[hbt!]
    \centering
    \includegraphics[width=0.9\linewidth]{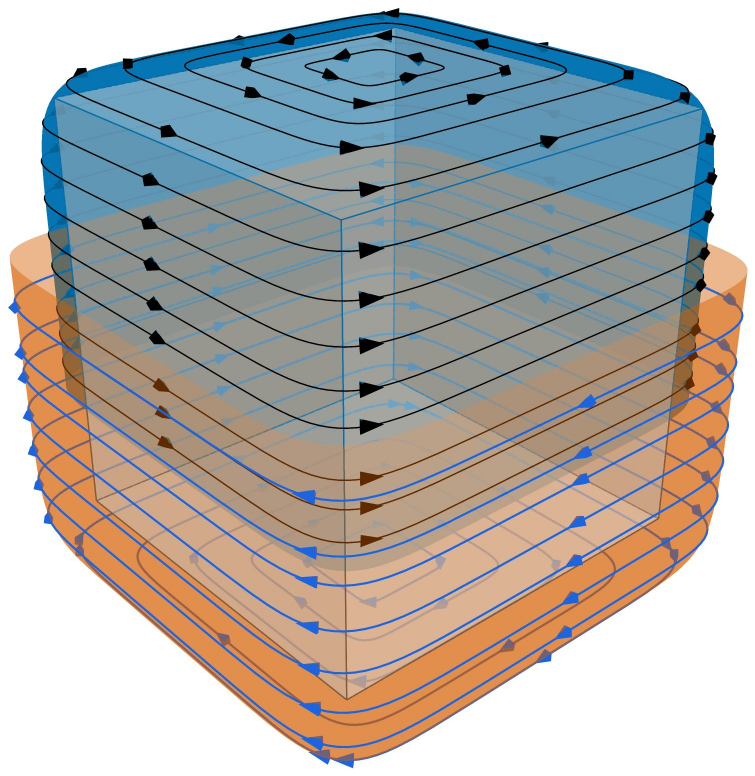}
    \caption{Two-patch Wu--Yang construction of the vector potential on the cube. Gauge  $\vec A_{\mathrm{I}}$ (blue region, black streamlines) covers $z>-z_0$, while gauge II  $\vec A_{\mathrm{II}}$ (orange region, blue streamlines) covers $z<z_0$; the unequal displayed extents are chosen only to make the overlap visible. Each vector potential produces the same outward face-normal magnetic field on its domain. In the overlap region, the two descriptions are related by the gauge transformation in Eq.~(\ref{eqn: Gauge Transformation}), and therefore give the same physical field and observables.
}
    \label{fig:gauge patches}
\end{figure}

The vector potential, rather than the magnetic field itself, enters the Schrödinger equation. The description is therefore gauge-dependent, although the spectrum and other physical quantities are not. For a magnetic monopole, this gauge freedom is not merely a convenience. Since the magnetic field has a nonzero total flux through a closed surface, it cannot be represented by a single smooth vector potential over the entire cube. We therefore describe the problem using two overlapping gauge patches, following the Wu--Yang construction for a monopole on the sphere \cite{wu_dirac_1976}. This choice is schematically represented in Fig.~(\ref{fig:gauge patches}).

We orient the cube so that its edges are parallel to the coordinate axes and its corners are at $(x,y,z)=(\pm \frac{1}{2},\pm \frac{1}{2},\pm \frac{1}{2})$. We choose a parameter $0<z_0<1/2$ and define two overlapping regions on the cube surface: region I consists of points with $z>-z_0$, while region II consists of points with $z<z_0$. To write the vector potentials compactly, we parametrize the boundary of a constant-$z$ cross section by a coordinate $s$, which is the periodic arclength coordinate with $0<s\leq 4$, chosen so that the curve runs counterclockwise around the square in the $xy$ plane:
\begin{equation}
(x(s),y(s))=
\begin{cases}
\left(s-\dfrac{1}{2},-\dfrac{1}{2}\right), & 0<s\leq 1,\\
\left(\dfrac{1}{2},s-\dfrac{3}{2}\right), & 1<s\leq 2,\\
\left(\dfrac{5}{2}-s,\dfrac{1}{2}\right), & 2<s\leq 3,\\
\left(-\dfrac{1}{2},\dfrac{7}{2}-s\right), & 3<s\leq 4.
\end{cases}
\end{equation}
Here, $z$ is held fixed throughout the parameterization.

The vector potential in the upper patch ($z>-z_0$) is defined as follows
\begin{equation}\label{eqn:A_I}
\vec{A}_\mathrm{I}(x,y,z)=
\begin{cases}
\dfrac{B x}{2}\,\hat{y}
-\dfrac{B y}{2}\,\hat{x},
& z=\dfrac{1}{2},\\[6pt]
\left[\left(\dfrac{1}{2}-z\right)B+\dfrac{B}{4}\right]\hat{s},
& -z_0<z<\dfrac{1}{2}.
\end{cases}
\end{equation}
This choice reduces to symmetric gauge on the top face and to the Landau gauge on the side faces, with the tangential components matched continuously across the upper edges.

Similarly, in the lower patch  ($z<z_0$) a second vector potential is defined
\begin{equation}\label{eqn:A_II}
\vec{A}_\mathrm{II}(x,y,z)=
\begin{cases}
-\dfrac{B x}{2}\,\hat{y}
+\dfrac{B y}{2}\,\hat{x},
& z=-\dfrac{1}{2},\\[6pt]
\left[\left(z+\dfrac{1}{2}\right)B+\dfrac{B}{4}\right](-\hat{s}),
& -\dfrac{1}{2}<z<z_0.
\end{cases}
\end{equation}

The two vector potentials describe the same magnetic field on the cube surface, and therefore must be related by a gauge transformation in the region where the two patches overlap. Since the overlap is the region $-z_0<z<z_0$, we write
\begin{equation}
\label{eqn: Gauge Transformation}
\vec{A}_{\mathrm{II}}=\vec{A}_{\mathrm{I}}-\vec{\nabla}\Lambda(\vec{r}).
\end{equation}
The wavefunctions in the overlap region are transformed accordingly:
\begin{equation}
\psi_{\mathrm{II}}(\vec r)
=
e^{-i\Lambda(\vec r)}
\psi_{\mathrm{I}}(\vec r).
\end{equation}
This gives the choice $\Lambda(\vec r)=\frac{3}{2}Bs$. Since $s$ parametrizes a closed curve with period 4, going once around the overlap region changes the gauge function by $6B$. The transition function relating the two wavefunctions must be single valued, which requires
\begin{equation}
e^{-i6B}=1.
\end{equation}
Therefore,
\begin{equation}
B=M\frac{\pi}{3},
\qquad M\in\mathbb{Z}.
\end{equation}
Complex conjugation maps the Hamiltonian at $M$ to that at $-M$, so the spectra are identical; henceforth we take $M \ge 0$.

Using Eq.~(\ref{eqn:flux quantum}), the total outward flux through the cube is therefore an integer multiple of the flux quantum,
\begin{equation}
\Phi_B=M\phi_0.
\end{equation}
This is the Dirac quantization condition in the present geometry \cite{dirac_quantised_1931,wu_dirac_1976}. It follows from requiring the transition function between the two gauge patches to be single-valued. Thus, the allowed monopole strengths are fixed by the global consistency of the quantum wavefunction.
This construction also shows why the symmetry analysis differs from the zero-field problem. The magnetic field is invariant under all rotations of the cube, but a particular choice of vector potential is not. As a result, ordinary rotation operators do not commute with the Hamiltonian written in a fixed gauge. In the next section, we construct the corresponding gauge-modified rotation operators. These operators provide the appropriate classification of the energy eigenstates in terms of cubic symmetry. 

\section{Symmetry and Gauge Transformations \label{sec:symmetry}}

\begin{table}[hbt]
\centering
\caption{Character table of the octahedral group $O$.}
\label{tab:O_chartable}
\renewcommand{\arraystretch}{1.3}
\setlength{\tabcolsep}{7pt}
\normalsize
\begin{tabular}{c|rrrrr}
\hline\hline
\rule{0pt}{2.8ex}%
$O$   & $E$ & $8C_3$ & $6C_4$ & $3C_2$ & $6C_2'$ \\
\hline
$A_1$ & $1$ & $1$    & $1$    & $1$    & $1$     \\
$A_2$ & $1$ & $1$    & $-1$   & $1$    & $-1$    \\
$E$   & $2$ & $-1$   & $0$    & $2$    & $0$     \\
$T_1$ & $3$ & $0$    & $1$    & $-1$   & $-1$    \\
$T_2$ & $3$ & $0$    & $-1$   & $-1$   & $1$     \\
\hline\hline
\end{tabular}
\end{table}
\begin{table}[hbt]
\centering
\caption{Character table of the binary octahedral group $2O$.}
\label{tab:2O_chartable}
\renewcommand{\arraystretch}{1.3}
\setlength{\tabcolsep}{3pt}
\normalsize
\begin{tabular}{c|rrrrrrrr}
\hline\hline
\rule{0pt}{2.8ex}%
$2O$ & $E$ & $\bar{E}$ & $8C_3$ & $8\bar{C}_3$ & $6C_4$ & $6\bar{C}_4$ & $6C_2$ & $12C_2'$ \\
\hline
$A_1$ & $1$ & $1$  & $1$  & $1$  & $1$        & $1$        & $1$  & $1$  \\
$A_2$ & $1$ & $1$  & $1$  & $1$  & $-1$       & $-1$       & $1$  & $-1$ \\
$E$   & $2$ & $2$  & $-1$ & $-1$ & $0$        & $0$        & $2$  & $0$  \\
$T_1$ & $3$ & $3$  & $0$  & $0$  & $1$        & $1$        & $-1$ & $-1$ \\
$T_2$ & $3$ & $3$  & $0$  & $0$  & $-1$       & $-1$       & $-1$ & $1$  \\
$G_1$ & $2$ & $-2$ & $1$  & $-1$ & $\sqrt{2}$ & $-\sqrt{2}$& $0$  & $0$  \\
$G_2$ & $2$ & $-2$ & $1$  & $-1$ & $-\sqrt{2}$& $\sqrt{2}$ & $0$  & $0$  \\
$H$   & $4$ & $-4$ & $-1$ & $1$  & $0$        & $0$        & $0$  & $0$  \\
\hline\hline
\end{tabular}
\end{table}
The surface of the cube is invariant under the rotational symmetry group of the cube, which is a subgroup of $O(3)$. The use of cubic symmetry to classify the eigenstates of the zero-field problem was developed in Ref.~\cite{cidlinsky_quantum_2024}. Throughout this work, we restrict our attention to  proper rotations and therefore use the cubic rotation group $O$. This group is of order 24, corresponding to the 4! permutations of 4 body-diagonals. The full octahedral group $O_h$, which is of order 48 as it includes the reflections and inversions, is not relevant for the magnetic Hamiltonian considered here, since the inversions switch the sign of the monopole charge. To be more precise, we will see below that for an odd monopole number, we will need the double cover, the binary octahedral group $2 O$ with 48 elements.  Although both $O_h$ and $2O$ have order $48$, they are distinct, non-isomorphic groups: $O_h \cong O \times \mathbb{Z}_2$ is the full octahedral point group, whereas $2O$ is the binary octahedral group, namely the double cover of the proper rotation group $O$ in $SU(2)$. Below, we expound on this more in the context of projective representations.

Each element of $O$ may be represented by a $3\times 3$ orthogonal matrix acting on the coordinates of the cube, together with a corresponding unitary operator acting on wavefunctions on the surface. As a concrete example, consider the operation of a $\pi/2$ rotation about the $z$ axis, passing through the center of the upper face of the cube. We call the operation $S$, and we can define two related operators. The first, $R_S$ is an orthogonal matrix rotating vectors in $\mathbb{R}^3$, e.g. $\vec{r}'=R_S \vec r$.  The other is the related operator in the Hilbert space of wavefunctions on the surface of the cube, ${\cal D}(R_S)$, which acts on the scalar wavefunctions in the active rotation sense:
\begin{equation}
{\cal D}(R_S) \Psi(\vec{r}) = \Psi(R_S^{-1} \vec{r}).
\end{equation}
The rotation operator is unitary ${\cal D}^\dagger(R_S)={\cal D}^{-1}(R_S)={\cal D}(R_S^{-1})$, as we have $R_S^{T}=R_S^{-1}$. The action of the rotation operator follows:
\begin{align}
    {\cal D}^\dagger(R_S) \,\vec{r} \,{\cal D}(R_S) = R_S \vec{r} \\
    {\cal D}^\dagger(R_S) \, \vec{p} \, {\cal D}(R_S) = R_S \vec{p} 
\end{align}

As noted above, the proper octahedral group $O$ consists of $24$ rotations, partitioned into five conjugacy classes: the identity; the six rotations by $\pi/2$ or $3\pi/2$ about axes through the centers of opposite faces; the three rotations by $\pi$ about the same axes; the eight rotations by $2\pi/3$ or $4\pi/3$ about axes through opposite vertices; and the six rotations by $\pi$ about axes through the midpoints of opposite edges. Correspondingly, $O$ has five inequivalent irreducible representations, conventionally denoted by $A_1$, $A_2$, $E$, $T_1$, and $T_2$, with dimensions $1$, $1$, $2$, $3$, and $3$, respectively. The character table used in our classification is given in Table~\ref{tab:O_chartable}. We adopt the standard notation and character conventions of Ref.~\cite{tinkham_group_2003}. In the numerical analysis below, the group is generated by a fourfold rotation about an axis through opposite faces and a threefold rotation about a body diagonal. These two rotations generate all $24$ elements of $O$. The characters of their matrix representations, restricted to each degenerate eigenspace, are then used to determine which irreducible representation is carried by that eigenspace.   

In the absence of a magnetic field, these rotation operators commute directly with the Hamiltonian. In the presence of a vector potential, however, the situation is more subtle because a spatial rotation transforms the vector potential as well as the coordinates. Thus, for the Hamiltonian Eq.~(\ref{eqn:dimensionless hamiltonian})
\begin{align}
    {\cal D}^\dagger(R_S) \,{\cal H} \,{\cal D}(R_S) &= -{\cal D}^\dagger(R_S) \,\left(\vec{\nabla}-i\vec{A}(\vec{r})\right)^2 \,{\cal D}(R_S) \nonumber \\ 
    &=-\left(R_S \vec{\nabla}-i\vec{A}(R_S\vec{r})\right)^2 \nonumber \\
    &=-\left(R_S \left (\vec{\nabla}-iR_S^{-1}\vec{A}(R_S\vec{r})\right)\right)^2 \nonumber \\
    &=-\left (\vec{\nabla}-iR_S^{-1}\vec{A}(R_S\vec{r})\right)^2.
\end{align}
So it would seem that the rotation operator, which is a symmetry of the cube, is a symmetry of the Hamiltonian only if
\begin{equation}
\vec{A}(\vec{r})=R_S^{-1}\vec{A}(R_S\vec{r}).
\end{equation}
This condition is overly restrictive because it requires the chosen vector potential to be rotationally invariant. The physical magnetic field may have the symmetry of the cube even when a particular gauge choice for the vector potential does not.

As in the spherical monopole problem, the resolution is to combine spatial rotations with gauge transformations \cite{wu_dirac_1976}. The same magnetic field can be described by different vector potentials related by a gauge transformation. To this end, we define the real valued gauge function $\Lambda(\vec{r})$ and the related operator
\begin{equation}
{\cal U}_\Lambda \Psi(\vec{r})=e^{i \Lambda(\vec{r})} \Psi(\vec{r}).
\end{equation}
This unitary operator is diagonal in real space but shifts momentum
\begin{align}
{\cal U}_\Lambda^\dagger \, \vec{r} \, {\cal U}_\Lambda &=\vec{r},  \nonumber \\
{\cal U}_\Lambda^\dagger \, \vec{\nabla} \, {\cal U}_\Lambda &=\vec{\nabla}+i \vec{\nabla}\Lambda(\vec{r})
\end{align}
We now define the combined rotation and gauge transformation operator:
\begin{equation}
\tilde{\cal D}(R_S,\Lambda_S)= {\cal D}(R_S){\cal U}_{\Lambda_S}.
\end{equation}
Applied to the Hamiltonian, this generates:
\begin{align}
\tilde{\cal D}^\dagger(R_S,\Lambda_S) &\,{\cal H} \,\tilde{\cal D}(R_S,\Lambda_S) \nonumber \\
= &-\left( \vec{\nabla} -i\left[R_S^{-1}\vec{A}(R_S\vec{r})-
\vec\nabla \Lambda_S(\vec{r})\right]\right)^2.
\end{align}
Thus, one can find a gauge transformation $\Lambda_S$ that  makes the combined operator commute with the Hamiltonian if there is a solution to
\begin{equation}
\vec{\nabla}\Lambda_S(\vec{r})= R_S^{-1}\vec{A}(R_S\vec{r})-\vec{A}(\vec{r}).
\end{equation}

This condition has a simple physical interpretation. Formally, the gauge function can be written as an integral along a path,
\begin{equation} \label{eqn:Gauge choice}
\Lambda_S(\vec{r})=\int_{\vec{r}_0}^{\vec{r}} d\vec{\ell'}\cdot(R_S^{-1}\vec{A}(R_S\vec{r'})-\vec{A}(\vec{r'})),
\end{equation}
but this expression is path independent only if 
\begin{equation}
\oint d\vec{\ell'}\cdot(R_S^{-1}\vec{A}(R_S\vec{r'})-\vec{A}(\vec{r'}))=0
\end{equation}
over any arbitrary closed path in our simply connected patches.  As the closed contour integral of the vector potential is the magnetic flux through the contour, this condition is equivalent to
\begin{equation}
\vec{B}(\vec{r})=R_S^{-1}\vec{B}(R_S\vec{r}),
\end{equation}
clearly showing that the gauge-adjusted rotation operator remains a symmetry if the magnetic field configuration is also symmetric under the said rotation. Thus, the apparent breaking of rotational symmetry by the vector potential is only a gauge artifact. The full rotational symmetry of the magnetic problem is restored by the gauge-modified operators $\tilde{\cal D}(R_S,\Lambda_S)$.

The construction above shows that the symmetry operations of the magnetic Hamiltonian are gauge-modified rotations rather than ordinary spatial rotations. For each rotation of the cube, one must accompany the coordinate transformation by a gauge transformation that restores the chosen vector potential to the original gauge. When the magnetic field itself has the full cubic symmetry, such a gauge transformation can be found, and the corresponding combined operator commutes with the Hamiltonian. Thus, the rotational symmetry of the physical magnetic field is recovered at the level of the quantum Hamiltonian, even though it may be hidden by a particular choice of vector potential. At the same time, Eq.~(\ref{eqn:Gauge choice}) determines the symmetry operator only up to an overall phase, reflected for example in the choice of the starting point for integration $\vec{r}_0$.  This phase freedom is not arbitrary: when two gauge-modified rotations are multiplied, the result may agree with the gauge-modified operator for the product rotation only up to a phase. The symmetry operators therefore need not form an ordinary representation of $O$, but may instead form a projective representation. Equivalently, the magnetic problem naturally leads to an extension of the cubic rotation group, a point that will be central when classifying the degeneracies of the numerical spectrum. Projective symmetry representations and their equivalent description in terms of double groups also arise in discrete monopole problems on polyhedral graphs \cite{oktel_spectrum_2012,kemp_discrete_2013}.

 The gauge-modified rotations introduced above are the rotational counterparts of magnetic translations \mbox{
\cite{brown_bloch_electrons,zak_magnetic_translation}}\hskip0pt
. In both constructions, a spatial symmetry of the magnetic field is accompanied by a compensating gauge transformation, so that the combined operator commutes with the Hamiltonian. The composition of these operators may introduce flux-dependent phases, leading to a projective representation of the spatial symmetry group. In the present problem, this gauge-induced phase structure is responsible for the distinction between ordinary and spinorial representations of cubic symmetry, even though the particle has no intrinsic spin.

This situation is the finite-symmetry analogue of the familiar Wu--Yang description of a particle moving in the field of a magnetic monopole. In that description, the wavefunction is defined patch by patch, and the transition functions between patches carry the information about the monopole charge. A rotation of the system must therefore be lifted not only to a transformation of the coordinates, but also to a transformation of the corresponding gauge bundle. For even monopole charge, the accumulated phases can be chosen so that the symmetry operations realize ordinary representations of the cubic rotation group. For odd monopole charge, however, a $2\pi$ rotation acts nontrivially on the wavefunction, and the relevant representations are double-valued representations of $O$. This is directly analogous to the double covering of $SO(3)$ by $SU(2)$, and motivates replacing $O$ by its double cover, the binary octahedral group $2O$.

The group $2O$ is the central extension of the cubic rotation group obtained by lifting the rotations of the cube to $SU(2)$ \cite{kemp_discrete_2013}. Equivalently, it is generated by the lifted versions of the fourfold face rotation and the threefold body-diagonal rotation introduced above, together with the central element corresponding to a $2\pi$ rotation. It contains 48 elements: each element of $O$ has two lifts that differ by this central element.
Its irreducible representations separate into two classes. The single-valued representations, for which the central element acts trivially, reduce to the ordinary irreducible representations $A_1$, $A_2$, $E$, $T_1$, and $T_2$ of the cubic rotation group. The character table of $2O$ used in the remainder of the paper is given in Table~\ref{tab:2O_chartable}. The spinorial representations,  $G_1$, $G_2$, and $H$, for which the central element acts as minus the identity, are required for odd monopole charge. In the remainder of the paper, we use the character table of $2O$ to classify the degenerate eigenspaces obtained numerically, with ordinary cubic representations appearing for even monopole charge and spinorial representations appearing for odd monopole charge. 

\section{Gauge-Covariant Numerical Method\label{sec:numerical}}

The numerical analysis consists of two steps. We first discretize the Hamiltonian on the surface of the cube using a gauge-covariant finite-difference scheme. Diagonalizing the resulting sparse Hermitian matrix gives the approximate energy levels and wavefunctions. We then construct the corresponding discrete symmetry operators, including the required gauge transformations, and use their traces within each degenerate manifold to assign irreducible representations of $O$ or $2O$.

We discretize each face of the cube by an $N_d\times N_d$ square lattice, identifying sites that lie on common edges and vertices of adjacent faces. This gives $N_{\mathrm{sites}}=6N_d^2-12N_d+8$ independent sites and $N_{\mathrm{plaq}}=6(N_d-1)^2$ elementary plaquettes. The wavefunction is represented by its values on these sites. The resulting lattice has four nearest neighbors at every site, except at the eight cube vertices, where only three links meet. With lattice spacing $\Delta x=1/(N_d-1)$, the continuum Laplacian is approximated by the usual nearest-neighbor finite-difference operator, so the graph-Laplacian eigenvalues are scaled by $(N_d-1)^2$ when calculating the continuum energies. Away from the cube vertices, this is the usual four-point stencil, while at the eight cube corners, the same graph-Laplacian construction uses the three links that meet there. For $M=0$, the scaled eigenvalues converge to the previously obtained spectrum of the zero-field cube \cite{cidlinsky_quantum_2024}.

The magnetic field is included through link variables rather than by assigning the vector potential directly to the lattice sites. For each nearest-neighbor bond $i\to j$, we define the phase factor
\begin{equation}
U_{ij}=\exp\left(i\int_i^j \vec{A}\cdot d\vec{\ell}\right),
\end{equation}
where the integral is evaluated along the link connecting the two sites. The reverse link has $U_{ji}=U_{ij}^{*}$. In the finite-difference Hamiltonian, each nearest-neighbor term is then multiplied by the corresponding link variable. This is the lattice version of minimal coupling: the Peierls description of particles moving on a lattice in a magnetic field \cite{hofstadter_energy_1976}. The wavefunction acquires the appropriate phase as it moves along a bond. The product of the link variables around an elementary plaquette gives the phase associated with the magnetic flux through that plaquette. A gauge transformation changes the phases assigned to individual links and the phases of the site-wavefunctions, but it leaves these plaquette products and the spectrum unchanged.

The vector potential is represented by two nonsingular gauge choices, defined on overlapping regions of the cube surface. In the numerical calculation, we choose a cut within this overlap region, typically the $z=0$ plane. Lattice sites above the cut are represented in gauge I, while sites below the cut are represented in gauge II. The cut is only a numerical convention, not a physical boundary.  The finite-difference equations are then written in the gauge assigned to the site at which the equation is evaluated. For sites far from the cut, all neighboring wavefunction values are already expressed in the same gauge. Only one additional step is needed for sites adjacent to the cut. For example, an equation written for a site just above the cut may contain the wavefunction at a neighboring site below the cut. This neighboring value is stored in gauge II, but the equation itself is written in gauge I, so it must first be converted using the transition function between the two gauges. Similarly, equations just below the cut use the inverse transformation for neighboring values above the cut. Equivalently, the hopping matrix elements crossing the cut are multiplied by the appropriate transition function, evaluated at the site whose wavefunction is being converted. Thus, all terms in each finite-difference equation are expressed in a single gauge, while neighboring equations on opposite sides of the cut are related by the Wu--Yang gauge transformation \cite{wu_dirac_1976}. This stitches the two patches together without imposing any extra boundary condition at the artificial cut.

\begin{figure}[hbt!]
    \centering
    \includegraphics[width=1\linewidth]{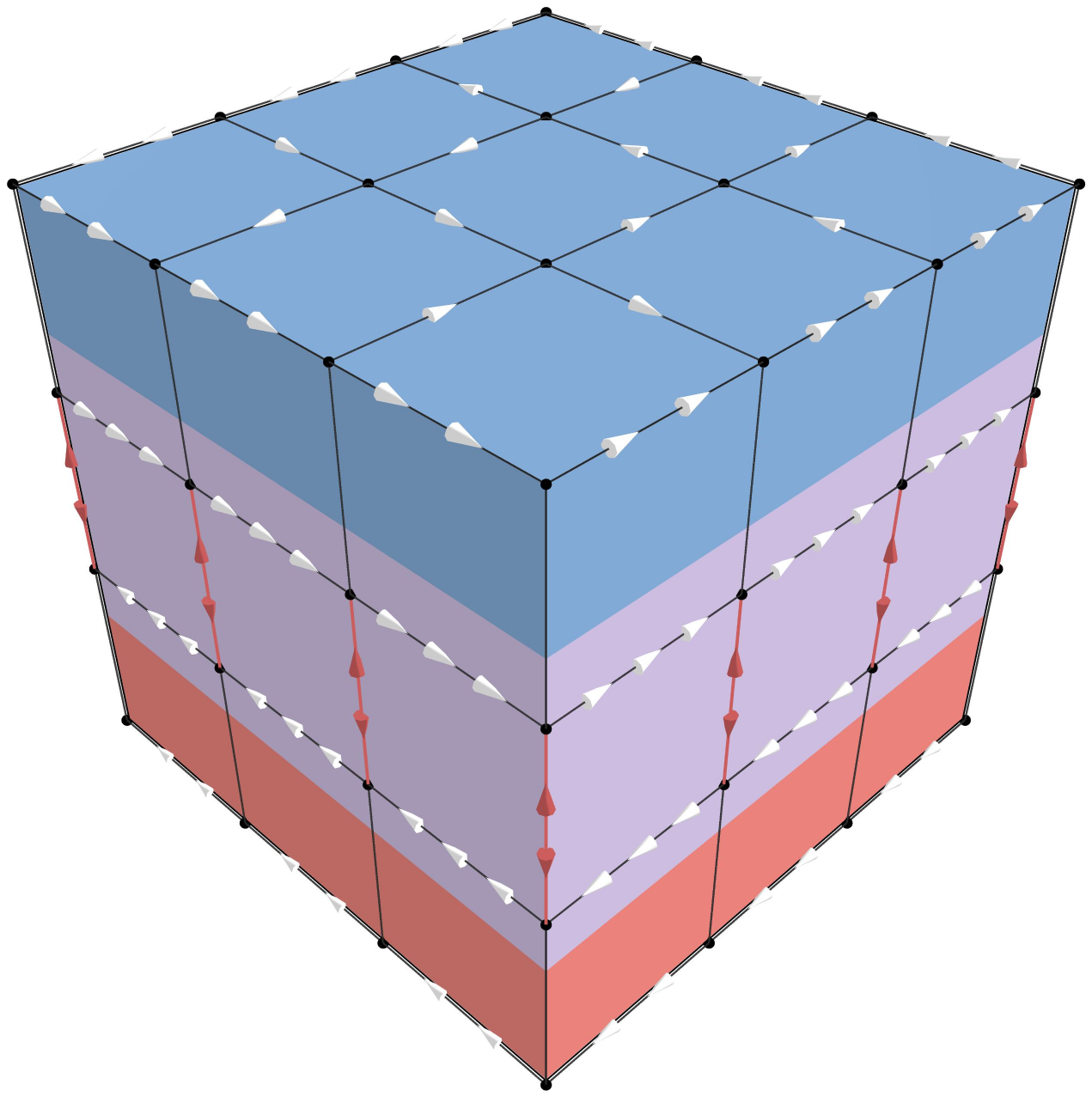}
    \caption{Gauge-covariant link variables and Wu--Yang patch stitching for $N_d=4$. Each face contains $(N_d-1)^2=9$ plaquettes, each carrying magnetic flux $\phi=B/9$. An arrow on a link indicates the direction in which its hopping phase is accumulated; reversing the hopping direction complex conjugates this phase. The finite-difference equations for sites with $z>0$ and $z<0$ are written in gauges I and II, respectively. Links crossing the artificial cut at $z=0$, shown by the red double arrows, contain the appropriate Wu--Yang transition factors. On the lateral faces their phase is proportional to $\Lambda(s)=3Bs/2$ and therefore increases linearly with the perimeter coordinate $s$. These stitching factors ensure that the oriented product of link variables around every plaquette, including those intersected by the cut, is $e^{i\phi}$.}
    \label{fig:lattice cube}
\end{figure}

Figure~\ref{fig:lattice cube} illustrates this construction for
$N_d=4$. Each face is divided into $(N_d-1)^2=9$ plaquettes, so the
magnetic flux through each plaquette is
$\phi=B/(N_d-1)^2=B/9$. An arrow on a link indicates the direction in
which the displayed hopping phase is accumulated; hopping in the
opposite direction carries the complex-conjugate phase. The link
variables in the upper and lower halves of the cube are calculated in
gauges I and II, respectively. For links crossing the artificial cut,
the hopping amplitude also contains the Wu--Yang transition factor.
On the lateral faces this factor depends on the perimeter coordinate
$s$ through $\Lambda(s)=3Bs/2$, and hence its phase varies linearly
along the cut. The two hopping directions carry mutually conjugate
transition factors, as indicated by the red double arrows in
Fig.~(\ref{fig:lattice cube}). With these factors included, the
oriented product of the link variables around every plaquette is
$e^{i\phi}$, including plaquettes intersected by the cut. The lattice
therefore carries a uniform magnetic flux without introducing a
physical seam between the two gauge patches.
The discretized Hamiltonian is assembled as a sparse Hermitian
matrix, and its low-lying eigenvalues and eigenvectors are obtained
using standard SciPy eigensolvers~\cite{virtanen_scipy_2020}.  Continuum energies are
estimated from the graph-Laplacian eigenvalues after multiplication by
$(N_d-1)^2$.  The individual low-energy levels converge sufficiently
rapidly with increasing $N_d$ to establish the spectral structure and
the energies at the precision displayed in Table~\ref{tab:spectrum_classification}.  The values
listed there were calculated at $N_d=18$ and should be understood as
representative finite-grid estimates rather than high-precision
continuum extrapolations.  

For the symmetry classification, we use the two rotations $C_4$ and $C_3$, shown in Fig.~(\ref{fig:cube and generators}), as generators. For even monopole numbers, these generate the cubic rotation group $O$, while for odd monopole numbers, they generate its double cover $2O$. On the lattice, each symmetry operation is represented by a rotation of the sites followed by the compensating gauge transformation Eq.~(\ref{eqn:Gauge choice}) in the previous section. This construction fixes the operator only up to an overall phase. We determine these phases by imposing the appropriate generator relations, such as $C_4^4=+1$ for ordinary representations and $C_4^4=-1$ for the spinorial case.

After diagonalization, nearly degenerate eigenvalues are grouped into manifolds using a fixed numerical tolerance. For each manifold, we restrict the discrete symmetry operators to the corresponding eigenspace and compute their traces. Comparing these characters with the character tables of $O$ or $2O$ gives the irreducible representation label of the manifold \cite{tinkham_group_2003}. The numerical implementation is available through the accompanying Zenodo repository \cite{Zenodo}, allowing the reader to obtain spectra for different values of $M$, higher-energy manifolds, and calculations at higher numerical precision.

\section{Results \label{sec:results}}

We calculate the energies and wavefunctions of the low-lying states and
classify each symmetry-protected multiplet according to the irreducible
representations of the relevant cubic symmetry group.
Table~\ref{tab:spectrum_classification} summarizes the energies,
degeneracies, and symmetry classifications of the lowest ten multiplets
for monopole numbers $M=0,\ldots,9$. For even $M$, the multiplets are
classified by irreducible representations of the cubic rotation group
$O$, while for odd $M$ they are classified by the spinorial irreducible
representations of the binary octahedral group $2O$. The tabulated
energies are finite-grid estimates obtained at $N_d=18$ and are quoted
only to the precision appropriate at this resolution. For $M=0$, our
results agree with the previously reported spectrum of a free particle
on the surface of a cube~\cite{cidlinsky_quantum_2024}, apart from Figure~14 
entries that we believe contain a shift of the labels. Representative
wavefunctions from selected multiplets are displayed in the appendix.

Figure~\ref{fig: low energy irreps} presents the energies listed in
Table~\ref{tab:spectrum_classification} and makes several features of the
spectrum more apparent. Although even and odd monopole numbers require
different symmetry classifications, the energies evolve continuously
with $M$. For odd $M$, every multiplet is either twofold or fourfold
degenerate, as required by the dimensions of the spinorial irreducible
representations of $2O$. The spectrum, which is broadly distributed at
small $M$, gradually organizes into bands separated by gaps as $M$
increases. For $M=0,\ldots,3$, the ground-state degeneracy is $M+1$, in
agreement with the Wu--Yang spectrum on the sphere~\cite{wu_dirac_1976}
and with the lowest-Landau-level degeneracy in Haldane's spherical
geometry~\cite{haldane_fractional_1983}. The case $M=4$ is more subtle.
The five states of the corresponding spherical lowest Landau level form
the $j=2$ representation, which branches under the cubic group as a
doublet and a triplet. At finite grid resolution these appear as two
distinct multiplets, with the triplet lying slightly lower in energy in
Table~\ref{tab:spectrum_classification}. However, their separation
decreases rapidly as $N_d$ is increased. Calculations up to $N_d=80$
show leading convergence close to $1/(N_d-1)^2$, while continuum
intercepts obtained from unconstrained fits drift systematically toward
zero as coarser grids are removed. We therefore do not resolve a
nonzero splitting in the continuum limit. Within our numerical
accuracy, the cubic doublet and triplet are consistent with an accidental
fivefold degeneracy, even though such a degeneracy is not required by
cubic symmetry. The splittings for higher $M$ values are robustly non-zero, 
showing the geometric broadening of the LLL. 

\begin{table*}[hbt!]
\centering
\small
\caption{Energies, degeneracies, and symmetry classifications of the
lowest ten energy manifolds for monopole numbers $M=0,\ldots,9$.
Each entry gives the energy on the first line and the corresponding
irreducible representation on the second line, with its degeneracy
shown in parentheses. For even $M$, the multiplets are classified by irreducible
representations of $O$, while odd $M$ are classified by spinorial
irreducible representations of $2O$. The results were obtained using
$N_d=18$.}
\label{tab:spectrum_classification}
\renewcommand{\arraystretch}{1.15}

\begin{tabular*}{\textwidth}{
    @{\extracolsep{\fill}}
    c
    *{10}{c}
}
\hline\hline
Manifold
& $M=0$ & $M=1$ & $M=2$ & $M=3$ & $M=4$
& $M=5$ & $M=6$ & $M=7$ & $M=8$ & $M=9$ \\[4pt]
\hline 
\\[-7pt]
1 & \shortstack{0.000\\$A_1\,(1)$} & \shortstack{1.045\\$G_1\,(2)$} & \shortstack{2.090\\$T_1\,(3)$} & \shortstack{3.133\\$H\,(4)$} & \shortstack{4.175\\$T_2\,(3)$} & \shortstack{5.215\\$G_2\,(2)$} & \shortstack{6.254\\$A_2\,(1)$} & \shortstack{7.293\\$G_2\,(2)$} & \shortstack{8.330\\$T_2\,(3)$} & \shortstack{9.366\\$H\,(4)$} \\[3pt]
2 & \shortstack{4.156\\$T_1\,(3)$} & \shortstack{7.218\\$H\,(4)$} & \shortstack{9.645\\$E\,(2)$} & \shortstack{12.619\\$H\,(4)$} & \shortstack{4.176\\$E\,(2)$} & \shortstack{5.217\\$H\,(4)$} & \shortstack{6.256\\$T_2\,(3)$} & \shortstack{7.296\\$H\,(4)$} & \shortstack{8.334\\$E\,(2)$} & \shortstack{9.373\\$H\,(4)$} \\[3pt]
3 & \shortstack{11.443\\$T_2\,(3)$} & \shortstack{15.855\\$G_2\,(2)$} & \shortstack{10.727\\$T_2\,(3)$} & \shortstack{15.098\\$G_2\,(2)$} & \shortstack{14.872\\$T_1\,(3)$} & \shortstack{16.871\\$G_1\,(2)$} & \shortstack{6.259\\$T_1\,(3)$} & \shortstack{7.301\\$G_1\,(2)$} & \shortstack{8.337\\$T_1\,(3)$} & \shortstack{9.382\\$G_1\,(2)$} \\[3pt]
4 & \shortstack{14.087\\$E\,(2)$} & \shortstack{18.514\\$H\,(4)$} & \shortstack{21.391\\$T_2\,(3)$} & \shortstack{26.194\\$H\,(4)$} & \shortstack{17.429\\$T_2\,(3)$} & \shortstack{19.852\\$H\,(4)$} & \shortstack{18.685\\$A_1\,(1)$} & \shortstack{21.791\\$G_1\,(2)$} & \shortstack{8.343\\$A_1\,(1)$} & \shortstack{27.991\\$H\,(4)$} \\[3pt]
5 & \shortstack{19.592\\$A_2\,(1)$} & \shortstack{29.505\\$G_2\,(2)$} & \shortstack{23.738\\$A_2\,(1)$} & \shortstack{29.901\\$G_1\,(2)$} & \shortstack{19.311\\$A_2\,(1)$} & \shortstack{22.368\\$G_2\,(2)$} & \shortstack{21.191\\$T_1\,(3)$} & \shortstack{25.403\\$H\,(4)$} & \shortstack{24.893\\$T_1\,(3)$} & \shortstack{32.982\\$G_2\,(2)$} \\[3pt]
6 & \shortstack{25.501\\$T_1\,(3)$} & \shortstack{32.438\\$G_1\,(2)$} & \shortstack{23.969\\$T_1\,(3)$} & \shortstack{29.979\\$G_2\,(2)$} & \shortstack{29.481\\$E\,(2)$} & \shortstack{35.151\\$H\,(4)$} & \shortstack{23.791\\$E\,(2)$} & \shortstack{28.734\\$H\,(4)$} & \shortstack{29.263\\$T_2\,(3)$} & \shortstack{34.143\\$H\,(4)$} \\[3pt]
7 & \shortstack{26.329\\$T_2\,(3)$} & \shortstack{33.401\\$H\,(4)$} & \shortstack{38.631\\$T_2\,(3)$} & \shortstack{44.005\\$H\,(4)$} & \shortstack{32.661\\$T_1\,(3)$} & \shortstack{38.971\\$G_1\,(2)$} & \shortstack{25.334\\$T_2\,(3)$} & \shortstack{44.284\\$G_2\,(2)$} & \shortstack{30.172\\$E\,(2)$} & \shortstack{36.521\\$G_1\,(2)$} \\[3pt]
8 & \shortstack{37.350\\$T_2\,(3)$} & \shortstack{47.265\\$H\,(4)$} & \shortstack{38.813\\$T_1\,(3)$} & \shortstack{47.205\\$H\,(4)$} & \shortstack{35.330\\$A_1\,(1)$} & \shortstack{41.648\\$H\,(4)$} & \shortstack{39.634\\$T_2\,(3)$} & \shortstack{45.456\\$H\,(4)$} & \shortstack{32.628\\$T_1\,(3)$} & \shortstack{52.831\\$H\,(4)$} \\[3pt]
9 & \shortstack{38.848\\$A_1\,(1)$} & \shortstack{50.173\\$G_1\,(2)$} & \shortstack{39.395\\$E\,(2)$} & \shortstack{49.316\\$G_1\,(2)$} & \shortstack{35.774\\$T_2\,(3)$} & \shortstack{52.090\\$G_2\,(2)$} & \shortstack{41.796\\$T_1\,(3)$} & \shortstack{51.990\\$G_1\,(2)$} & \shortstack{48.915\\$T_2\,(3)$} & \shortstack{55.990\\$G_2\,(2)$} \\[3pt]
10 & \shortstack{44.396\\$T_1\,(3)$} & \shortstack{54.588\\$H\,(4)$} & \shortstack{42.171\\$A_1\,(1)$} & \shortstack{66.927\\$H\,(4)$} & \shortstack{48.340\\$T_2\,(3)$} & \shortstack{60.418\\$H\,(4)$} & \shortstack{47.058\\$T_1\,(3)$} & \shortstack{52.729\\$H\,(4)$} & \shortstack{49.569\\$E\,(2)$} & \shortstack{61.501\\$H\,(4)$} \\[3pt]
\hline\hline
\end{tabular*}
\end{table*}

Figure~\ref{fig: Landau levels} shows the spectrum up to $M=30$. On this energy scale, the small splittings are no longer resolved, making the emerging Landau-level structure more apparent. In our units, the continuum Landau-level energies are
\begin{equation}
E_n= \frac{\hbar \omega_c (n+1/2)}{\hbar^2/(2mL^2)}=\frac{\pi}{3}(2n+1)M,    
\end{equation}
and the corresponding lines for the lowest three Landau levels are included in the figure. As $M$ increases, an increasing number of states cluster around these lines. In particular, the broadened lowest Landau level contains $M+1$ states, while its width remains small compared with the gap to the first excited Landau level. A distinct set of states also appears in the gap between the first and second excited Landau levels. Unlike a Landau level, whose number of states grows linearly with $M$, this intermediate manifold always contains eight states. These states originate from the eight corners of the cube, which act as geometric defects and support localized in-gap modes. We examine this phenomenon in the discrete lattice problem in the next section, where its geometric origin becomes more transparent.

\begin{figure}[hbt!]
    \centering
    \includegraphics[width=1\linewidth]{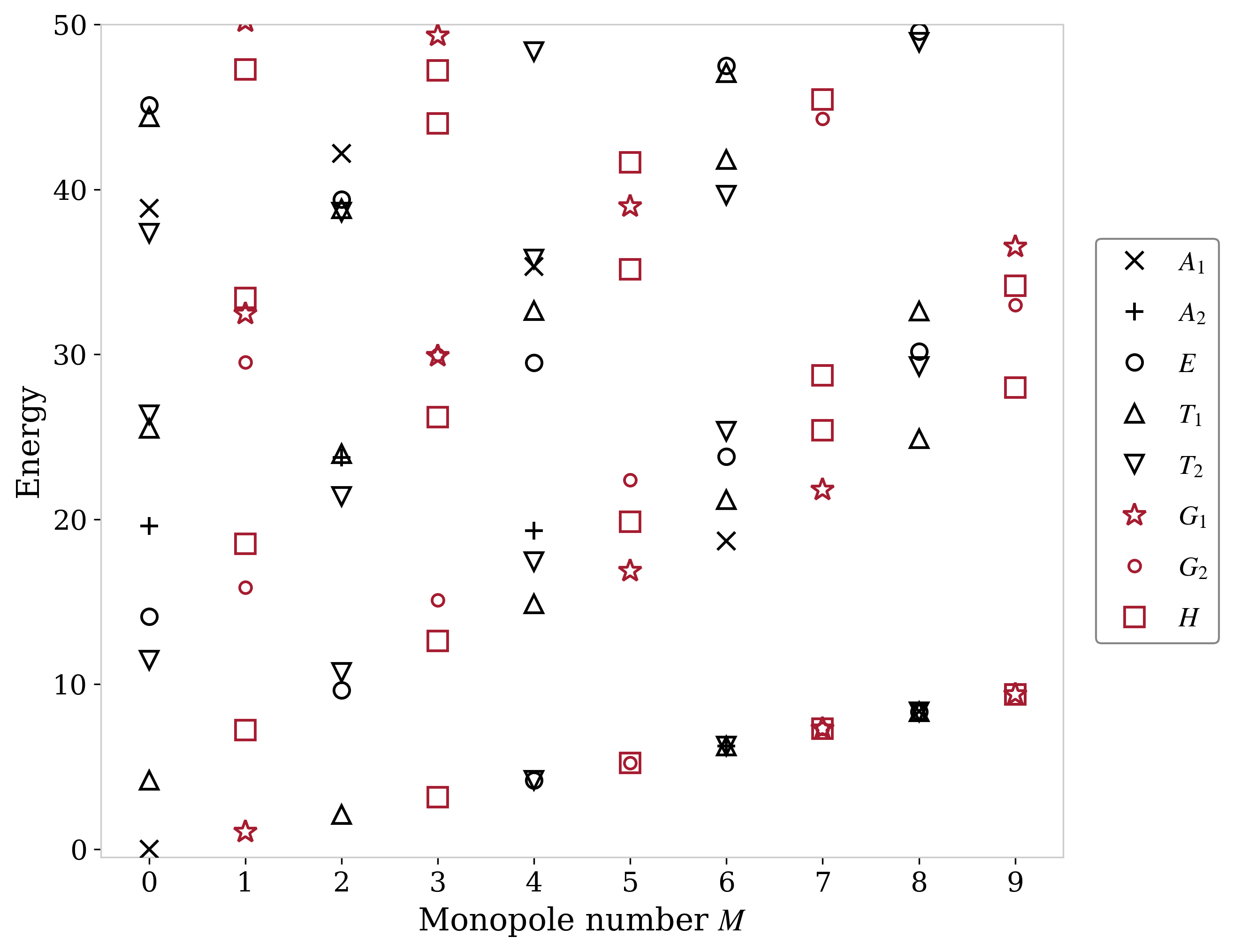}
    \caption{Energy spectrum of the ten lowest manifolds as a function of monopole number $M$, computed at $N_d=18$ and listed in Table~\ref{tab:spectrum_classification}. Manifolds are classified by the irreducible representations of the cubic rotation group $O$ for even $M$, or by the spinorial representations of the binary octahedral group $2O$ for odd $M$.}
    \label{fig: low energy irreps}
\end{figure}
\begin{table}[t]
    \caption{
    Branching of the continuum Wu--Yang LLL multiplet \(D^{(j)}\),
    with \(j=M/2\), under the rotational symmetry of the cube. For
    even \(M\), the multiplet decomposes into ordinary irreducible
    representations of \(O\); for odd \(M\), it decomposes into the
    spinorial irreducible representations \(G_1\), \(G_2\), and \(H\)
    of \(2O\). The dimensions on the right-hand side add to the
    continuum LLL degeneracy \(M+1\). Repeated irreps indicate
    multiplicities and need not remain at the same energy after the
    cubic splitting. 
    }
    \label{tab:LLL-branching}
    \centering
    \begin{ruledtabular}
        \begin{tabular}{ccl}
            \(M\) & \(j=M/2\) & Cubic decomposition \\[2pt]
            \hline
             0 & \(0\)             & \(A_1\) \\
             1 & \(1/2\)           & \(G_1\) \\
             2 & \(1\)             & \(T_1\) \\
             3 & \(3/2\)           & \(H\) \\
             4 & \(2\)             & \(E\oplus T_2\) \\
             5 & \(5/2\)           & \(G_2\oplus H\) \\
             6 & \(3\)             & \(A_2\oplus T_1\oplus T_2\) \\
             7 & \(7/2\)           & \(G_1\oplus G_2\oplus H\) \\
             8 & \(4\)             & \(A_1\oplus E\oplus T_1\oplus T_2\) \\
             9 & \(9/2\)           & \(G_1\oplus 2H\) \\
            10 & \(5\)             & \(E\oplus 2T_1\oplus T_2\) \\
            11 & \(11/2\)          & \(G_1\oplus G_2\oplus 2H\) \\
            12 & \(6\)             & \(A_1\oplus A_2\oplus E
                                        \oplus T_1\oplus 2T_2\) \\
        \end{tabular}
    \end{ruledtabular}
\end{table}
\begin{figure}[hbt!]
    \centering
    \includegraphics[width=1\linewidth]{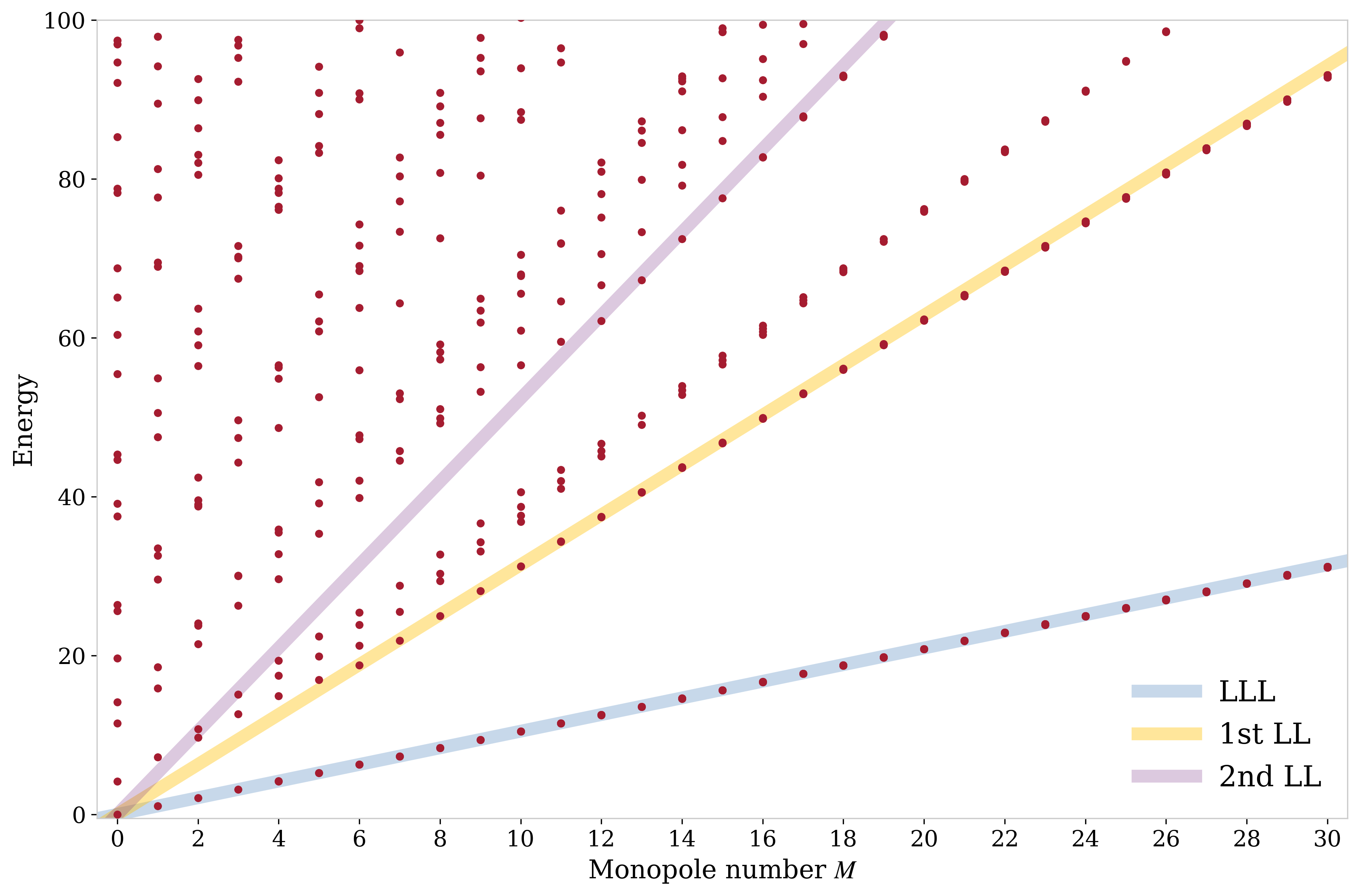}
    \caption{Energy spectrum up to $M=30$ at $N_d=18$. Dashed lines show the continuum
Landau-level energies $E_n=\frac{\pi}{3}(2n+1)M$  for $n=0,1,2$.
The broadened lowest Landau level contains $M+1$ states; the eight states in the gap between the $n=1$ and $n=2$ levels consist of the
localized states further discussed in Sec.~VI.} 
    \label{fig: Landau levels}
\end{figure}

\begin{figure*}[t]
    \centering
    \includegraphics[width=0.95\textwidth]
    {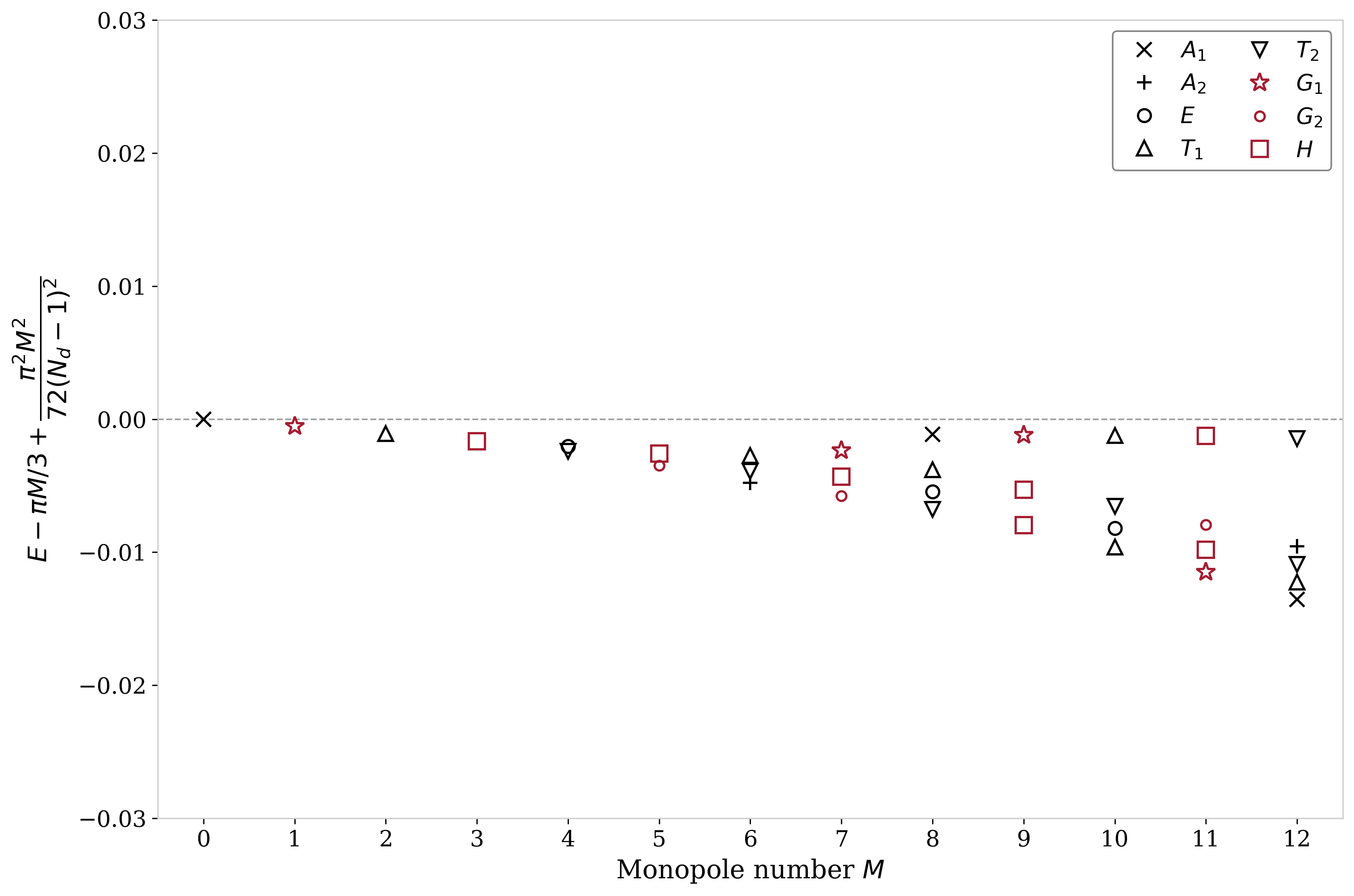}
    \caption{
    Splitting of the lowest Landau level (LLL) by the cubic geometry for
    monopole numbers \(M=0,\ldots,12\). Each symbol represents a degenerate
    energy manifold and identifies its irreducible representation of \(O\)
    for even \(M\), or its spinorial irreducible representation of \(2O\)
    for odd \(M\). The resulting decompositions agree with the group-theory
    branching rules for the \(j=M/2\) multiplet listed in
    Table~\ref{tab:LLL-branching}. The continuum LLL energy
    \(E_{\mathrm{LLL}}=\pi M/3\) has been subtracted, together with the
    leading common finite-difference correction. The results were obtained
    using \(N_d=28\) and provide our best numerical estimate of the
    continuum LLL splittings. Since the dominant discretization error is
    approximately common to all states within a given LLL, the relative
    splittings are expected to be more accurate than their absolute
    positions relative to \(E_{\mathrm{LLL}}\).
    }
    \label{fig:LLL-splitting}
\end{figure*}

The Wu--Yang monopole harmonics provide the natural continuum
reference for the symmetry content of the lowest Landau level
\cite{wu_dirac_1976}. For monopole number \(M\geq 0\), the monopole
harmonics are organized into angular-momentum multiplets with
\(j=M/2+n\), where \(n\) is the Landau-level index. The LLL therefore
corresponds to the multiplet \(D^{(j)}\) with \(j=M/2\) and has
dimension \(2j+1=M+1\). On the cube, the continuous rotational
symmetry is reduced to the proper cubic rotation group. The expected
splitting is consequently obtained by restricting the continuum
multiplet \(D^{(j)}\) to \(O\) for integer \(j\), corresponding to even
\(M\), and to its double cover \(2O\) for half-integer \(j\),
corresponding to odd \(M\). The character of a rotation through an
angle \(\theta\) in the angular-momentum-\(j\) representation is
\cite{tinkham_group_2003}
\begin{equation}
    \chi_j(\theta)
    =
    \frac{\sin[(j+1/2)\theta]}{\sin(\theta/2)}.
\end{equation}
The multiplicity of an irrep \(\Gamma\) in the restricted multiplet is
then found from the character inner product
\begin{equation}
    n_{\Gamma}^{(j)}
    =
    \frac{1}{|G|}
    \sum_{\mathcal C}
    |\mathcal C|\,
    \chi_{\Gamma}^{*}(\mathcal C)
    \chi_j(\theta_{\mathcal C}),
    \qquad
    G=O\ \text{or}\ 2O,
\end{equation}
where the sum runs over the conjugacy classes of the appropriate
group. In the spinorial case, \(\theta_{\mathcal C}\) denotes the
angle associated with the corresponding lift to \(SU(2)\). Applying
this reduction gives the branching rules listed in
Table~\ref{tab:LLL-branching}.

The numerical results provide a direct confirmation of these branching
rules. For every monopole number \(M=0,\ldots,12\), the degeneracies and
irreducible representations of the LLL manifolds displayed in
Fig.~(\ref{fig:LLL-splitting}) agree exactly with the decomposition listed
in Table~\ref{tab:LLL-branching}, including the multiplicities of repeated
irreducible representations.

Finally, in the appendix, we present a gallery of representative symmetry-adapted wavefunctions from the low-lying manifolds. We focus on the values of $M=0,1,4,5$ to provide two examples of even and odd monopole numbers each. The Hamiltonian and the fourfold rotation $C_4$ about a cube face play roles analogous to those of $L^2$ and $L_z$ in a rotationally invariant problem. We therefore diagonalize $C_4$ within each degenerate energy manifold, thereby resolving the degeneracy. For even $M$, the possible $C_4$ eigenvalues are the four roots of unity, whereas for odd $M$ they are the four roots of $-1$, reflecting the spinorial character of the corresponding states. Each wavefunction is displayed both on the cube, where we plot $|\Psi|^2$, and on a planar unfolding of its surface, where domain coloring also reveals the phase. The Wu--Yang patching is clearly visible in the phase plots, while the probability density shows no artificial distinction between the northern and southern patches, providing a useful validation of the numerical implementation. As the energy increases, the wavefunctions develop more nodes and show an increasing tendency toward vortex-like phase winding around the cube corners.

\begin{figure*}[hbt!]
    \centering
    \includegraphics[width=1\linewidth]{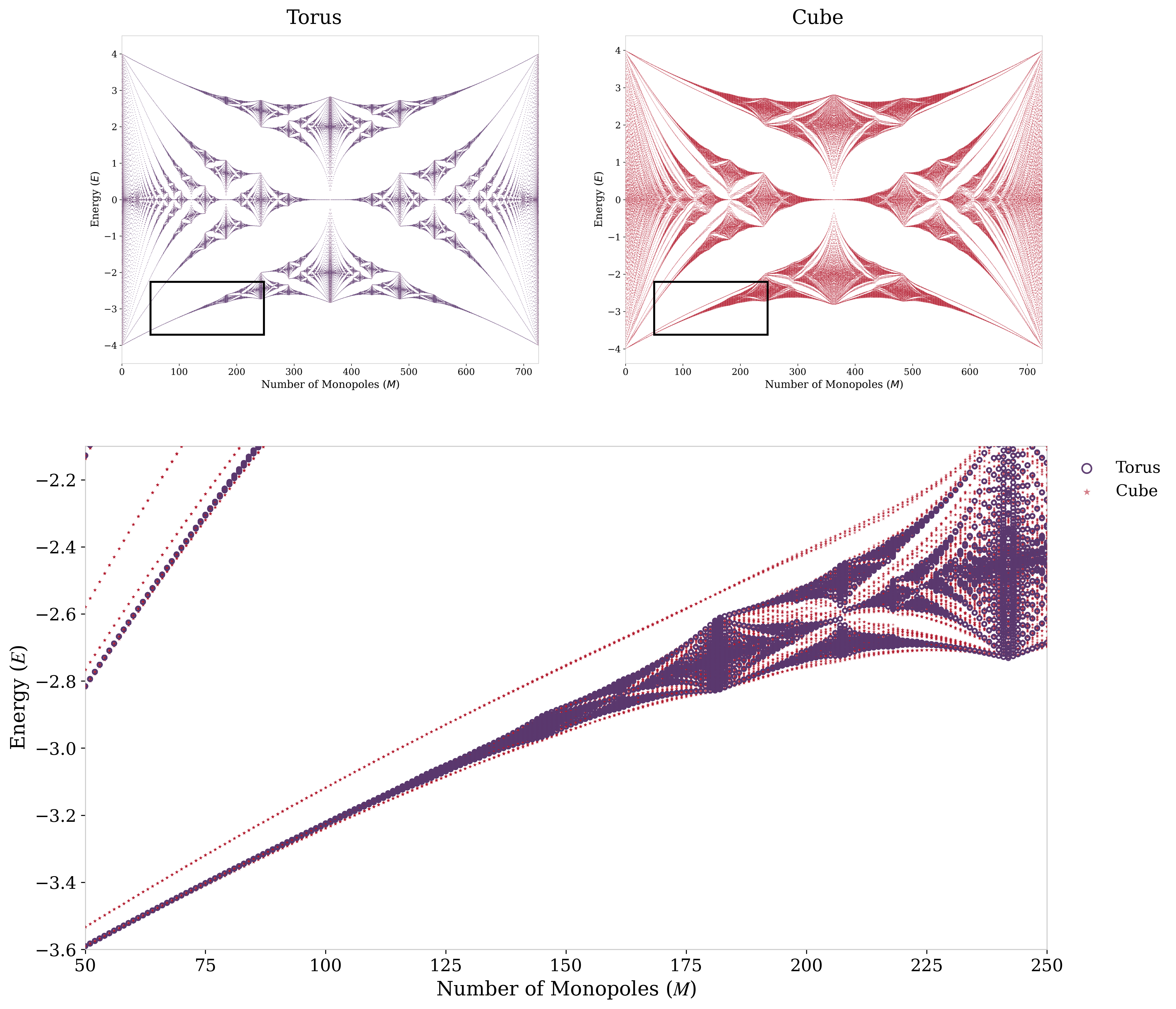}
    \caption{Comparison of the Hofstadter spectra of an $N_d=12$ cube and a $22\times33$ torus. The cube has 726 plaquettes and 728 sites, while the torus has 726 sites. In both calculations, the diagonal onsite terms are omitted, leaving the pure magnetic hopping Hamiltonian. Top: complete spectra for the torus (purple) and cube (red); the boxes mark the low-$M$ interval enlarged below. Bottom: overlay of the corresponding low-$M$ spectra. Gaps that remain empty on the torus contain additional cube states, identified as corner-localized states.}
    \label{fig: torus vs no diagonal}
\end{figure*}

\section{Hofstadter Butterfly on the cube \label{sec:butterfly}}

\begin{figure}[hbt!]
    \centering
    \includegraphics[width=1\linewidth]{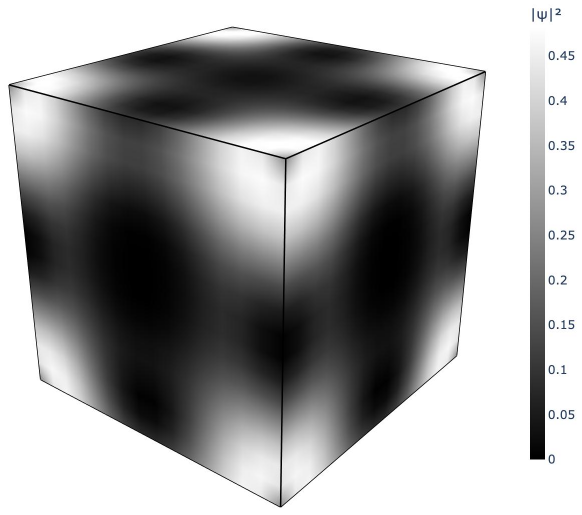}
    \caption{Probability density $|\Psi|^2$ of a representative state from a $T_2$ gap multiplet for $M=24$ with $N_d=16$. The state has $E= -3.77, \lambda = i$, clearly lying above the LLL value, which spans roughly the interval $[-4,4]$.  The density is strongly concentrated near the eight cube vertices, demonstrating the corner-localized character of the additional gap states. Their localization originates from the threefold, rather than fourfold, coordination of the corner sites.}
\label{fig: corner state}
\end{figure}

\begin{figure}[hbt!]
    \centering
    \includegraphics[width=1\linewidth]{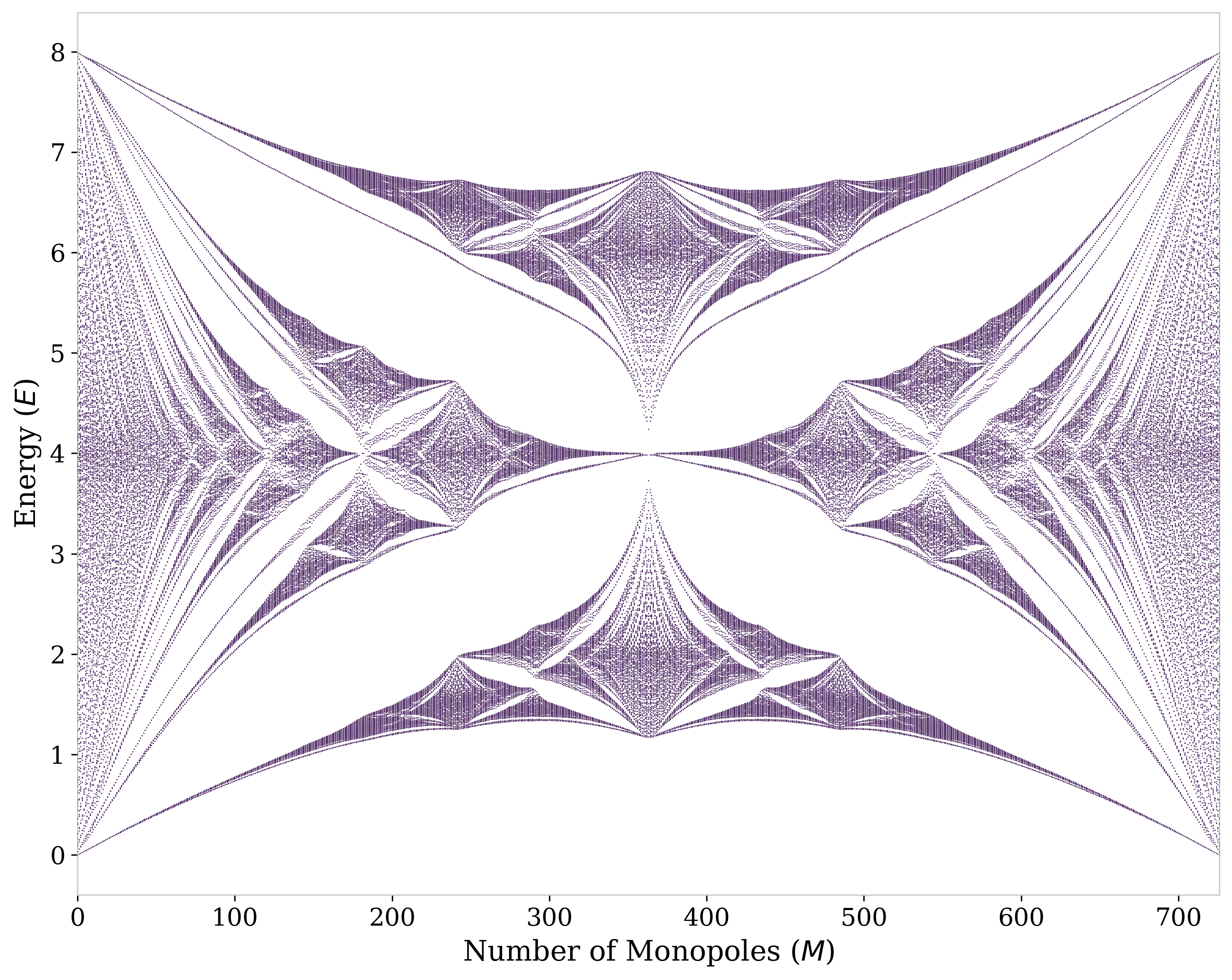}
    \caption{Hofstadter spectrum of the $N_d=12$ cube with the diagonal terms of the graph Laplacian restored: onsite energy 4 at ordinary sites and 3 at the eight corner sites. These terms move the lowest corner-localized gap states into the broadened lowest band. They also break the bipartite symmetry, removing the exact $E\rightarrow-E$ symmetry of the pure hopping Hamiltonian shown in Fig.~(\ref{fig: torus vs no diagonal}).}
    \label{fig: cube laplacian}
\end{figure}

The numerical spectra discussed above were obtained by discretizing the continuum Hamiltonian on the surface of the cube. The same discretized surface, however, also defines a natural tight-binding problem in its own right. We place an $N_d\times N_d$ square lattice on each face and identify sites on common edges and vertices as before. The Hilbert space then consists of one orbital on each independent lattice site. The magnetic field enters through Peierls phases on nearest-neighbor hopping terms. Starting from the gauge-covariant finite-difference Hamiltonian, the tight-binding Hofstadter model \cite{hofstadter_energy_1976} is obtained by dropping the diagonal onsite terms and keeping only the magnetic hopping matrix.   

Magnetic tight-binding models on graphs formed by the vertices of regular polyhedra have been studied previously \cite{avishai_tight-binding_2008,oktel_spectrum_2012,kemp_discrete_2013}. In contrast, the present lattice covers the entire surface of each cube face, and its number of sites increases with $N_d$. It therefore retains the local structure of the square-lattice Hofstadter problem while also incorporating the global geometry of the cube.

We compute the tight-binding spectrum as a function of the number of flux quanta through the cube. The result is shown in Fig.~(\ref{fig: torus vs no diagonal}). The spectrum displays the familiar splitting of the square-lattice Hofstadter problem into many magnetic subbands, separated by gaps that evolve continuously with flux. Thus, the closed-cube surface retains the basic magnetic interference mechanism of the Hofstadter problem, even though translational symmetry is replaced by the finite rotational symmetry of the cube. Related connections between continuum Landau levels and Hofstadter bands, including the use of long-range hopping to recover Landau-level degeneracies on a lattice, were investigated in Ref.~\cite{atakisi_landau_2013}.

To separate the effects of the local square lattice from those of the cube geometry, we compare this spectrum with the usual Hofstadter butterfly on a square lattice with periodic boundary conditions, shown in Fig.~(\ref{fig: torus vs no diagonal}). The overall structure is similar in the two cases. However, the cube spectrum in Fig.~(\ref{fig: torus vs no diagonal}) contains additional states inside several of the gaps. These states are absent in the torus calculation and therefore cannot be attributed to the local square-lattice Hofstadter physics alone.

The spatial structure of these gap states reveals their geometric origin. A representative example is shown in Fig.~(\ref{fig: corner state}). The probability density is strongly localized near the corners of the cube. The number of such gap states is always eight, independent of the lattice size $N_d$, matching the number of cube corners. These states arise from the special coordination of the corner sites, where only three links meet instead of four. They are therefore not edge states in the usual planar-boundary sense, but corner states tied to the singular geometry of the polyhedral surface.

The role of singular vertices in Landau-level spectra
has also been investigated through an analytical calculation
on a cone and a tight-binding Hofstadter model on a cube
\mbox{
\cite{biswas_son_fractional_charge}}\hskip0pt
. This work distinguished curvature-induced
inter-Landau-level states from additional modes
sensitive to the local structure near the vertices.

We then restore the onsite terms inherited from the finite-difference Laplacian. These terms assign diagonal energy $4$ to ordinary sites and $3$ to the eight corner sites, according to the number of nearest neighbors. The resulting spectrum is shown in Fig.~(\ref{fig: cube laplacian}). With these onsite terms included, the corner-localized states in the lowest-band gap are pushed back into the broadened lowest band. This effect was already apparent in the continuum calculation in Fig.~(\ref{fig: Landau levels}).  The gap states are sensitive to the distinction between the adjacency Hamiltonian and the graph-Laplacian Hamiltonian. At the same time, the onsite terms break the bipartite particle-hole symmetry of the pure nearest-neighbor hopping model, so the spectrum no longer has the exact $E\rightarrow -E$ symmetry seen in Fig.~(\ref{fig: torus vs no diagonal}). 

 This comparison gives a useful consistency check on the lattice construction. The broad Hofstadter pattern follows from the local square-lattice hopping, while the additional gap states in Fig.~(\ref{fig: corner state}) are tied to the corners of the closed cube surface. Restoring the graph-Laplacian diagonal terms, as in Fig.~(\ref{fig: cube laplacian}), brings the model back closer to the finite-difference Hamiltonian used for the continuum problem. The Hofstadter calculation therefore separates the local lattice physics from effects caused by the polyhedral geometry.

\section{Conclusions \label{sec:conclusion}}

We have studied the quantum mechanics of a charged particle confined to the surface of a cube enclosing a magnetic monopole. The model combines locally free motion on each face with a gauge structure that is global in character. Using a magnetic field of constant magnitude normal to each face, we constructed a two-patch description of the vector potential and showed that consistency of the wavefunction gives the Dirac quantization condition in this geometry.

The magnetic field preserves the rotational symmetry of the cube, but any explicit choice of vector potential hides this symmetry. The appropriate symmetry operations are therefore rotations supplemented by gauge transformations. These gauge-modified rotations commute with the Hamiltonian and provide the basis for classifying the eigenstates. Even monopole charges are described by the ordinary irreducible representations of the cubic rotation group $O$, while odd monopole charges require the spinorial representations of the binary octahedral group $2O$.

The resulting spectra show Landau-level-like manifolds on the cube. These manifolds are not degenerate in the same way as on the sphere, since the continuous rotational symmetry is reduced to the discrete symmetry of the cube. The low-energy degeneracies are therefore determined by the irreducible representations of $O$ or $2O$. Symmetry-adapted wavefunctions give a direct visualization of this structure and show how the nodal patterns and phases are organized by cubic rotations.

We also considered the corresponding tight-binding Hofstadter problem on the discretized cube surface. The spectrum contains the familiar magnetic subband structure, but comparison with a torus geometry shows additional states in several gaps. These states are localized near the eight cube vertices and originate from the reduced coordination of the corner sites. Restoring the onsite terms of the graph Laplacian moves the lowest corner states back into the broadened lowest band and removes the particle-hole symmetry of the pure hopping model.

Several extensions are natural. The same construction can be
applied to other polyhedral surfaces, where different point
groups and vertex structures should lead to different
splittings and localized states. The connection with compact
quantum Hall geometries
\cite{haldane_fractional_1983,kudo_crossover_2024}
may also be useful for finite-size studies in the presence
of interactions.

Such studies could explore the competition between
interactions and the possible geometric splitting of the
$M+1$ states in the lowest-Landau-level manifold, as well
as the local charge distribution near the cube vertices.
Boundary conditions have been shown to affect spectra
and Hall response in a strip geometry
\mbox{
\cite{angelone2023boundary}}\hskip0pt
. Although the cube has no physical boundary, the role
of local matching conditions at its singular vertices
remains an interesting question.

Finally, established techniques for producing synthetic
gauge fields \cite{goldman_light-induced_2014} and
advances in light-shaped optical confinement of
ultracold gases \cite{navon_quantum_2021} provide
ingredients for exploring related geometrically
structured models.

\acknowledgments
A. Kura was supported by a Bilkent Undergraduate scholarship during part of this work.
The scripts and data generated for this work are available on Zenodo \cite{Zenodo}.

\appendix*
\section{Symmetry-adapted wavefunction gallery}
\label{app:gallery}
The figures in this appendix present representative symmetry-adapted
wavefunctions for the even monopole numbers $M=0,4$ and the odd
monopole numbers $M=1,5$. Within each degenerate multiplet, we
diagonalize the gauge-modified fourfold rotation $C_4$ and label the
resulting states by its eigenvalue $\lambda$. For even $M$, the allowed
eigenvalues are the four roots of unity, whereas for odd $M$ they are
the four roots of $-1$. Each state is shown both through its
probability density $|\Psi|^2$ on the closed cube and through a planar
unfolding in which domain coloring encodes the phase $\arg(\Psi)$.
Although the Wu--Yang transition between gauge patches is visible in
the phase, the probability density remains continuous across the
numerical cut. All displayed wavefunctions are obtained numerically with $N_d=16$.

\begin{figure*}[hbt!]
    \centering
    \includegraphics[width=1\linewidth]{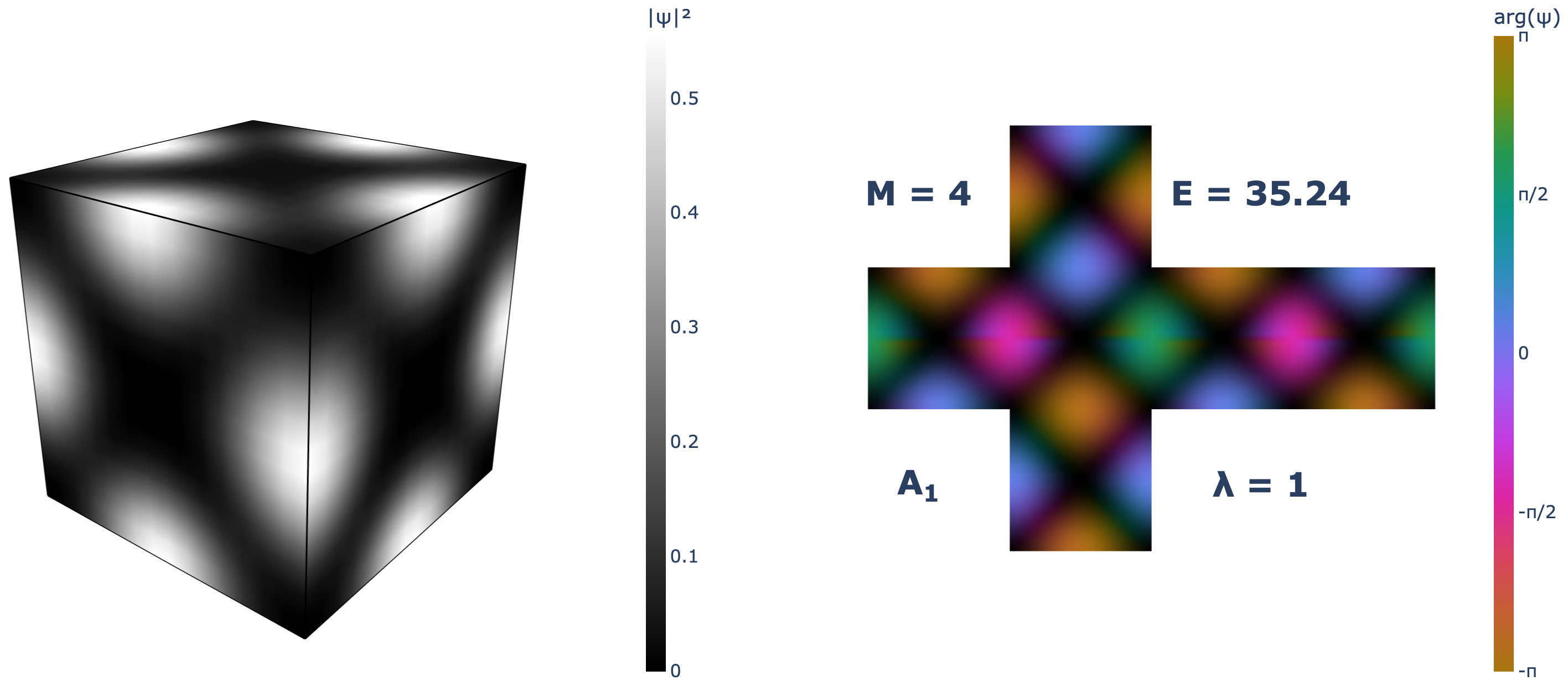}
    \caption{Visualization conventions used in the wavefunction gallery, illustrated with an $A_1$ state at $M=4$ and $(E=35.24,\lambda=1)$. Left: probability density $|\Psi|^2$ on the closed cube, displayed in grayscale. Right: planar unfolding of the same state, with hue encoding $\arg(\Psi)$ from $-\pi$ to $\pi$. These density and phase conventions are used in all subsequent figures.}
    \label{fig: colorbar}
\end{figure*}

\begin{figure*}[hbt!]
    \centering
    \begin{minipage}[t]{0.49\textwidth}
        \centering
        \includegraphics[width=\linewidth]{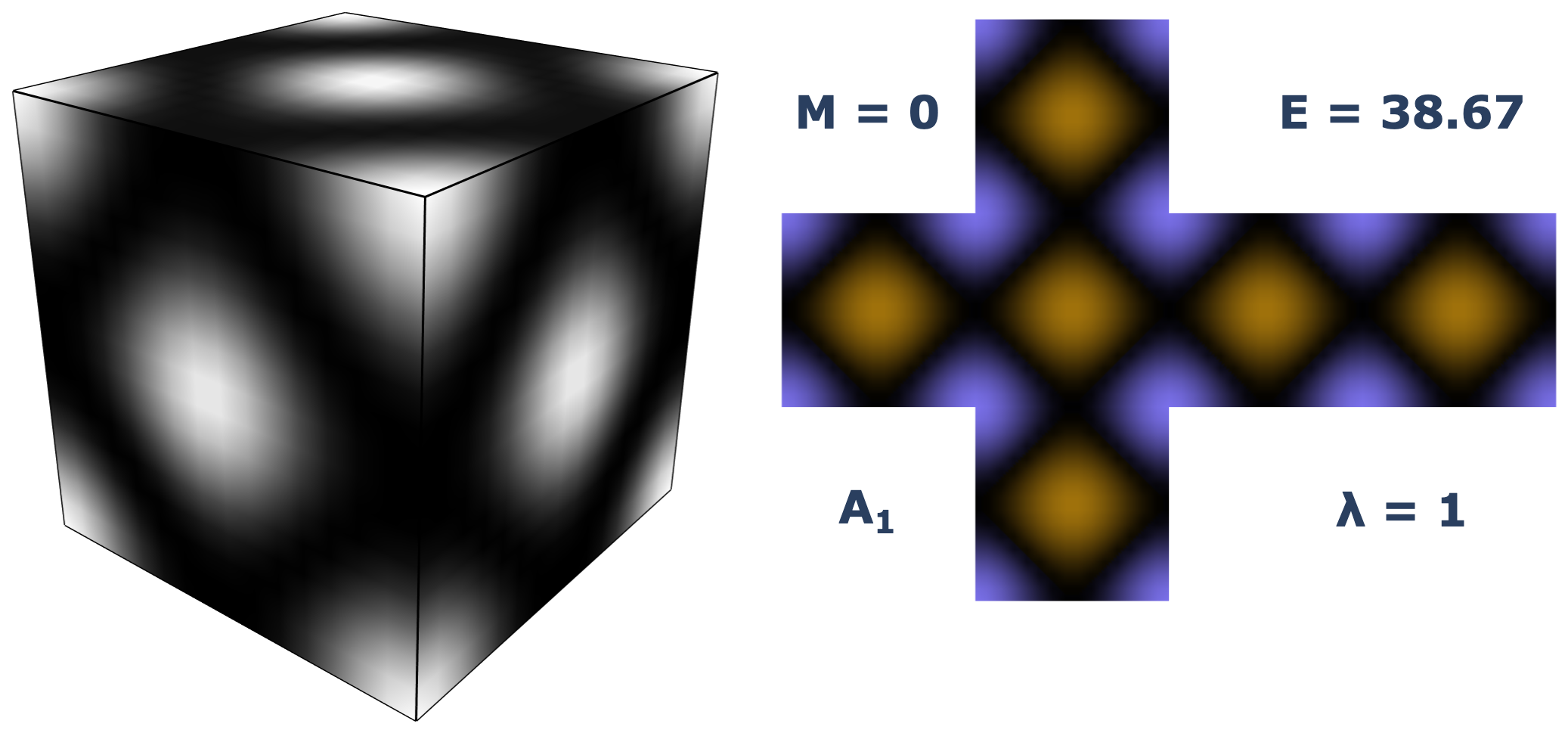}
    \end{minipage}
    \hfill
    \begin{minipage}[t]{0.49\textwidth}
        \centering
        \includegraphics[width=\linewidth]{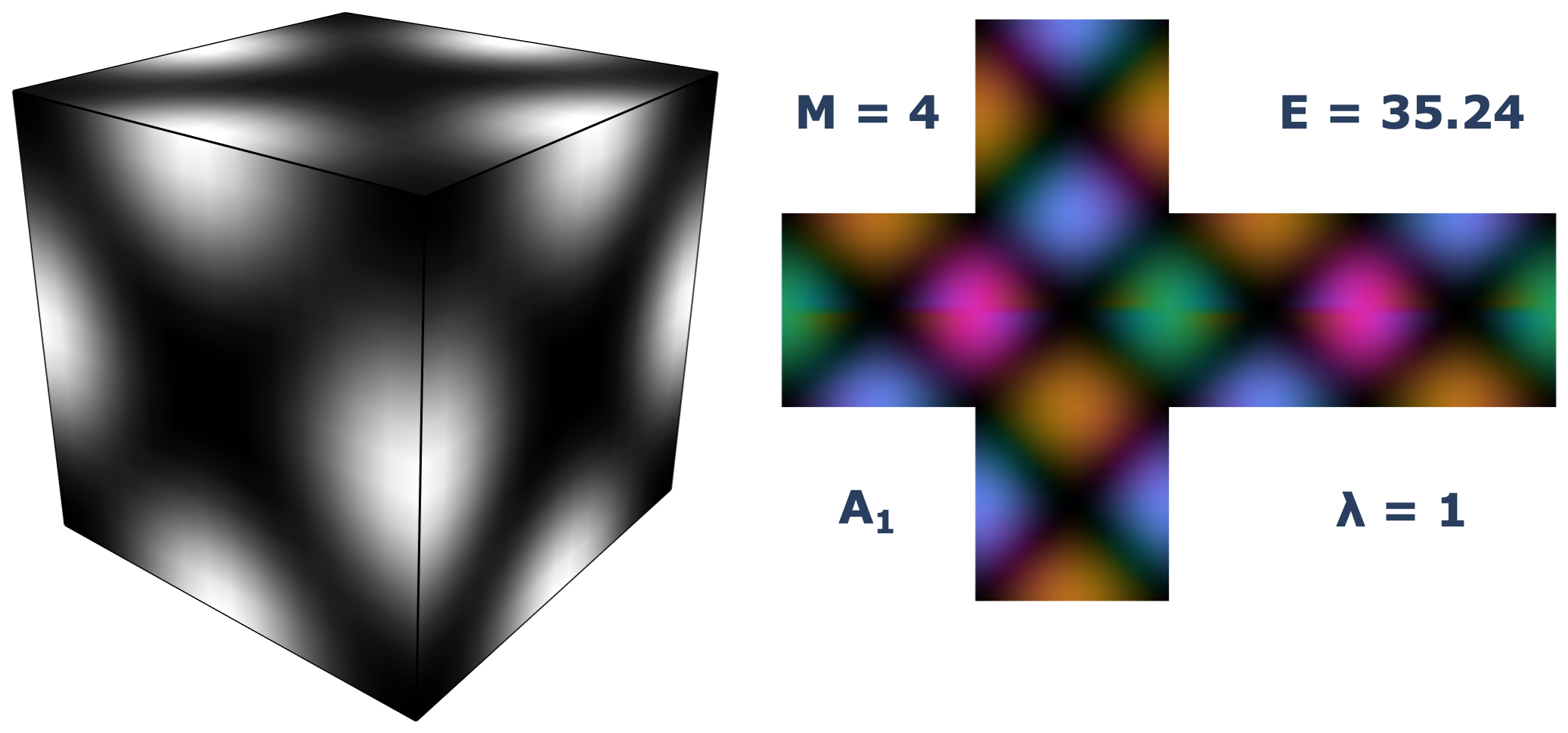}
    \end{minipage}
    \caption{Second lowest $A_1$ state for $M=0$ (left) and lowest $A_1$ state for $M=4$ (right). The lowest $A_1$ state for $M=0$ is uniform on the cube, thus not displayed here.}
    \label{fig: M0_M4_A1}
\end{figure*}

Figure~\ref{fig: colorbar} defines the common
color conventions. Figures~\ref{fig: M0_M4_A1}--%
\ref{fig: M1_M5_G2} then organize selected states by irreducible
representation, displaying a complete $C_4$ eigenbasis for each
multiplet. The final four figures provide a broader survey: one
representative $C_4$ eigenstate is selected from each of the eight
lowest energy multiplets for $M=0,1,4,$ and $5$, ordered by increasing
energy. As the energy increases, the probability densities develop
additional nodes, while the phase plots exhibit increasingly intricate
winding patterns, particularly near the edges and corners of the cube.

\begin{figure*}[hbt!]
    \centering
    \begin{minipage}[t]{0.49\textwidth}
        \centering
        \includegraphics[width=\linewidth]{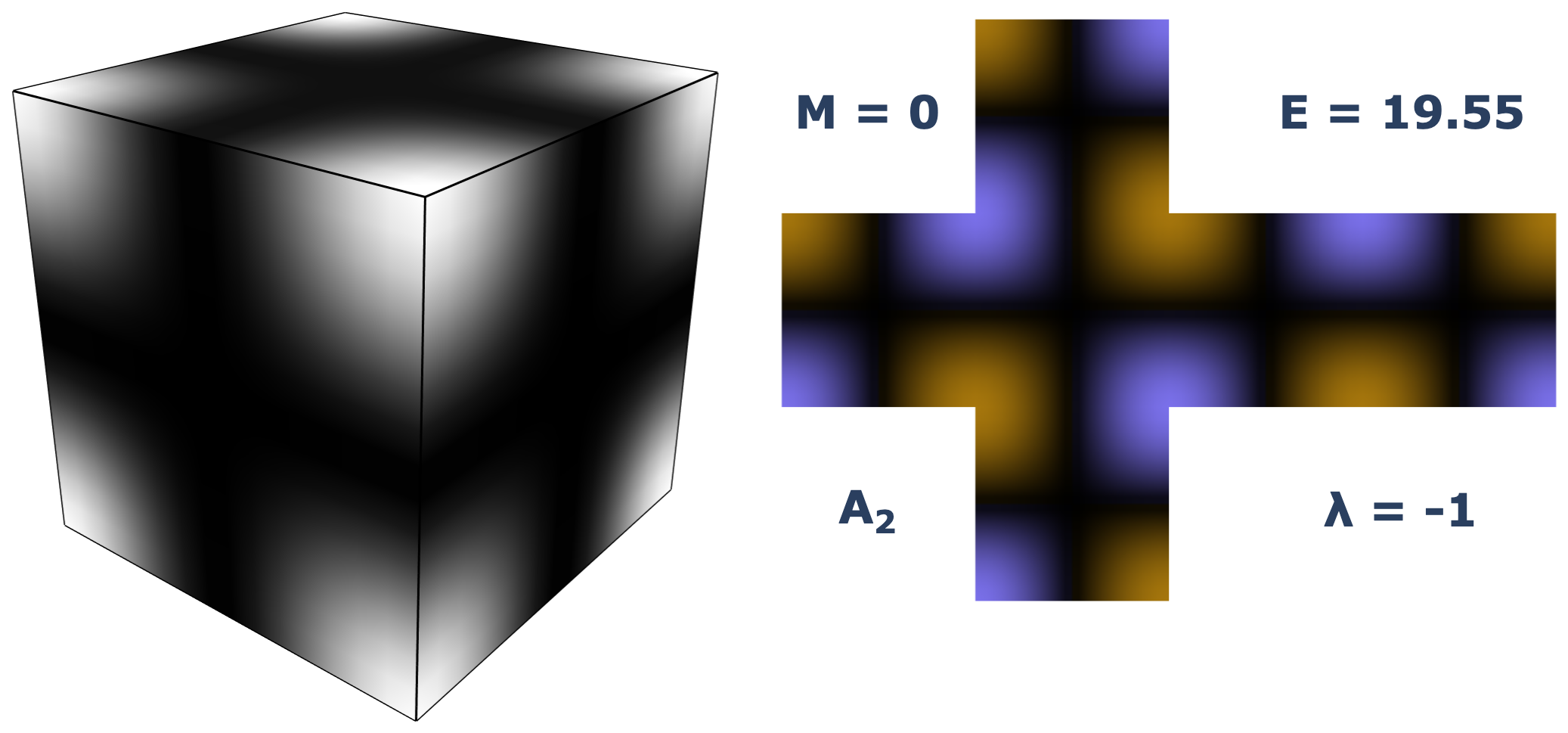}
    \end{minipage}
    \hfill
    \begin{minipage}[t]{0.49\textwidth}
        \centering
        \includegraphics[width=\linewidth]{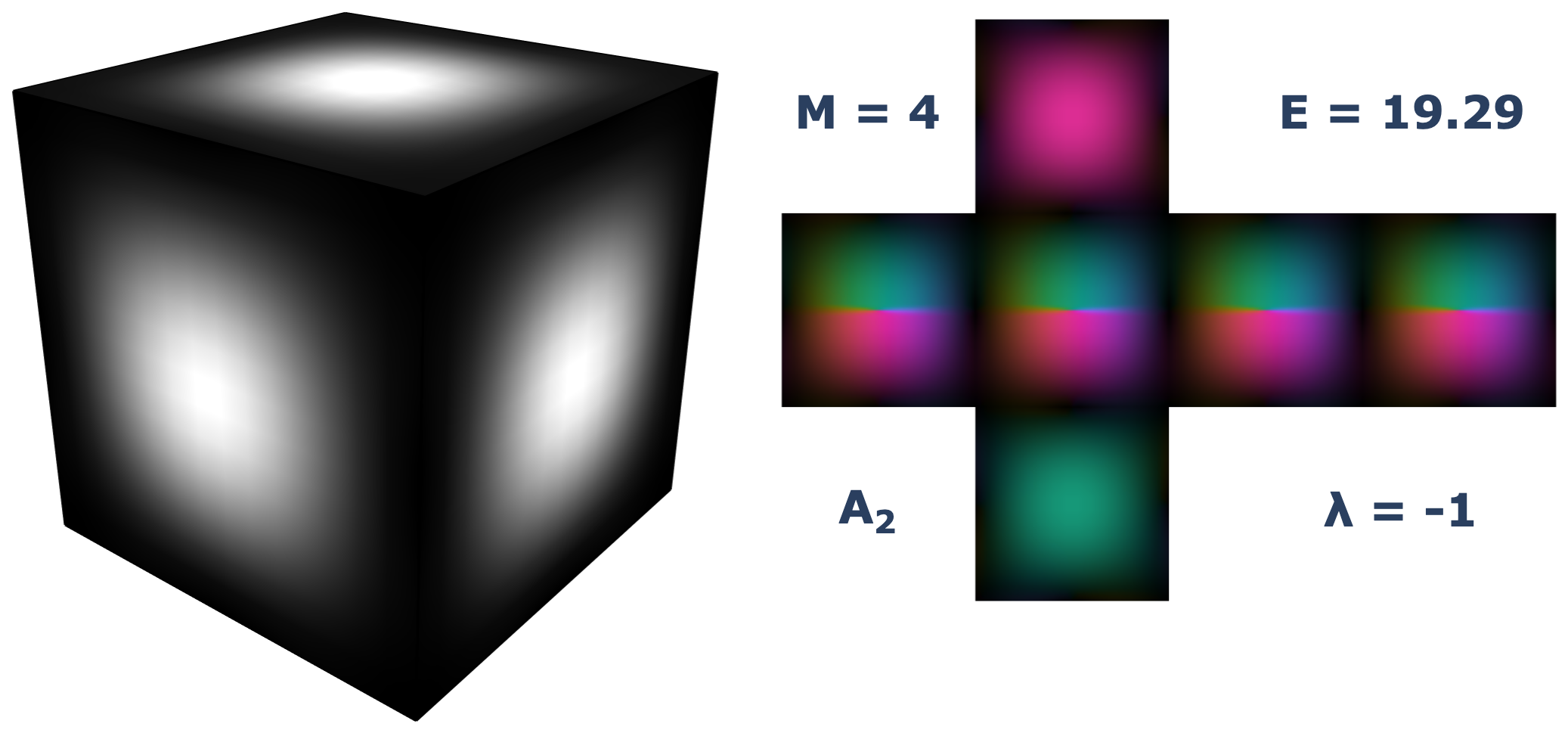}
    \end{minipage}
    \caption{Lowest $A_2$ state for $M=0$ (left) and $M=4$ (right).}
    \label{fig: M0_M4_A2}
\end{figure*}
\begin{figure*}[hbt!]
    \centering
    \begin{minipage}[t]{0.49\textwidth}
        \centering
        \includegraphics[width=\linewidth]{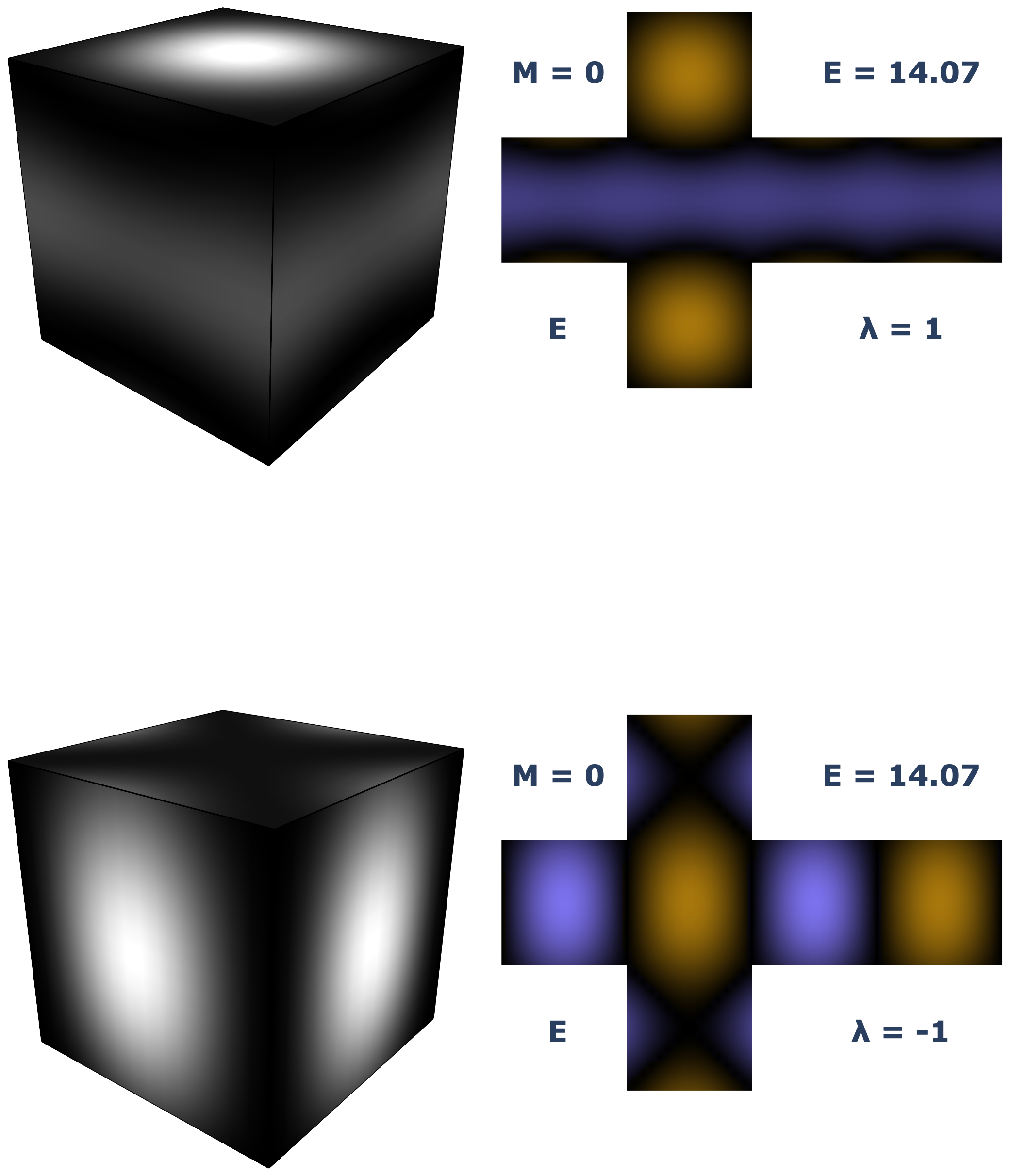}
    \end{minipage}
    \hfill
    \begin{minipage}[t]{0.49\textwidth}
        \centering
        \includegraphics[width=\linewidth]{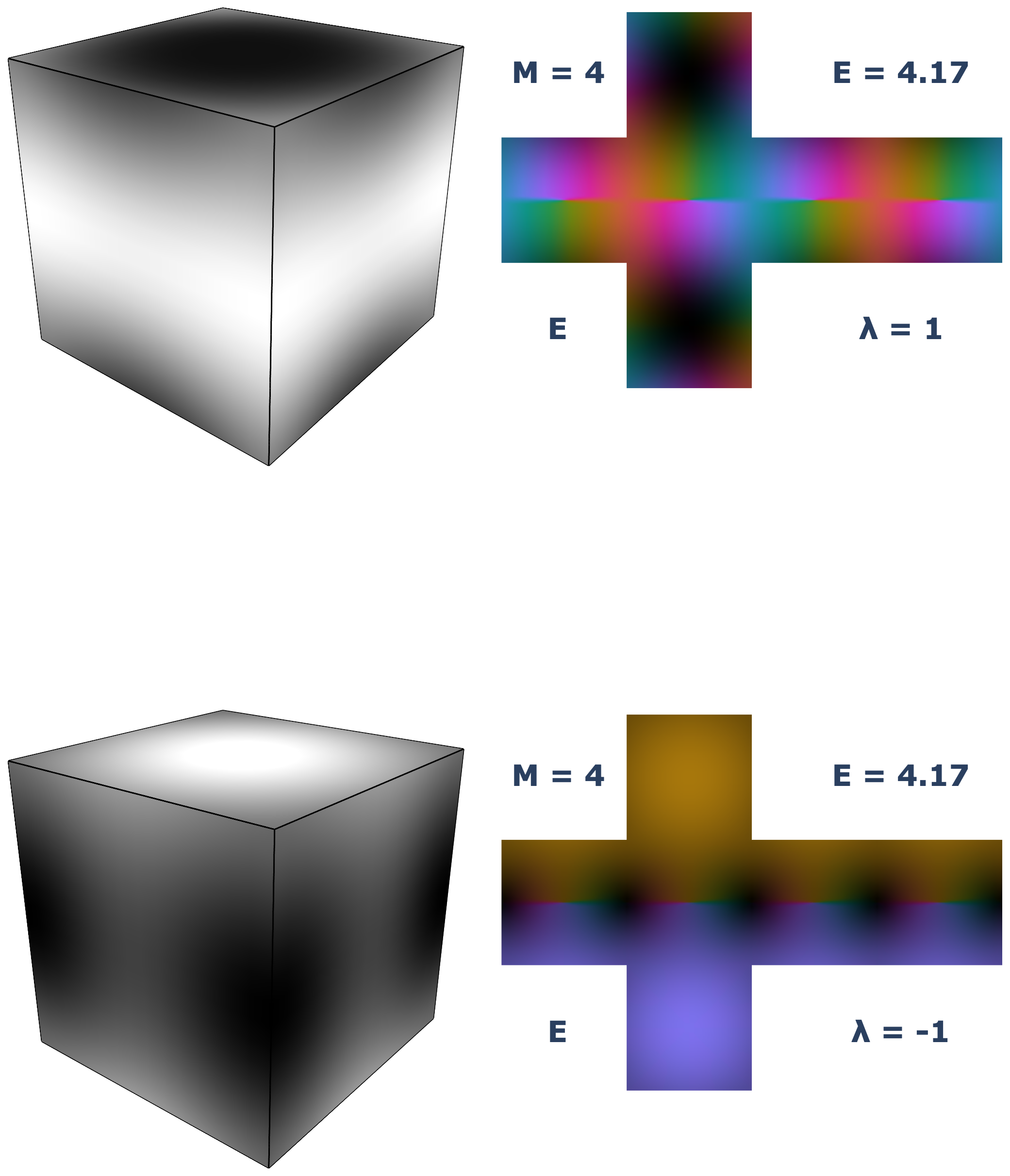}
    \end{minipage}
    \caption{Lowest $E$ multiplet at $N_d=16$ for $M=0$ (left) and $M=4$ (right). The rows display the complete $C_4$  eigenbasis of the multiplet; the corresponding $\lambda$ values are printed in the panels. }
    \label{fig: M0_M4_E}
\end{figure*}
\begin{figure*}[hbt!]
    \centering
    \begin{minipage}[t]{0.49\textwidth}
        \centering
        \includegraphics[width=\linewidth]{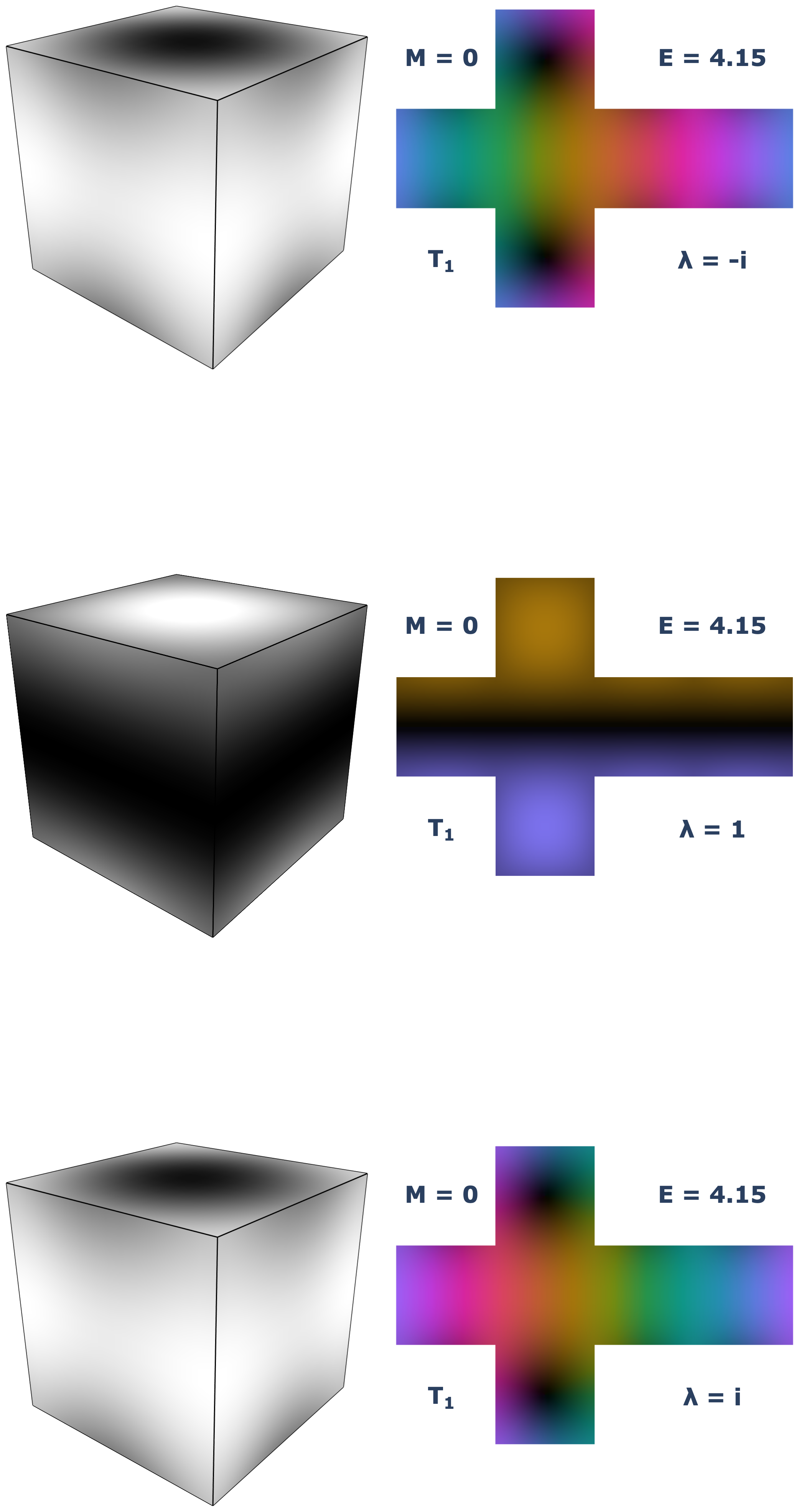}
    \end{minipage}
    \hfill
    \begin{minipage}[t]{0.49\textwidth}
        \centering
        \includegraphics[width=\linewidth]{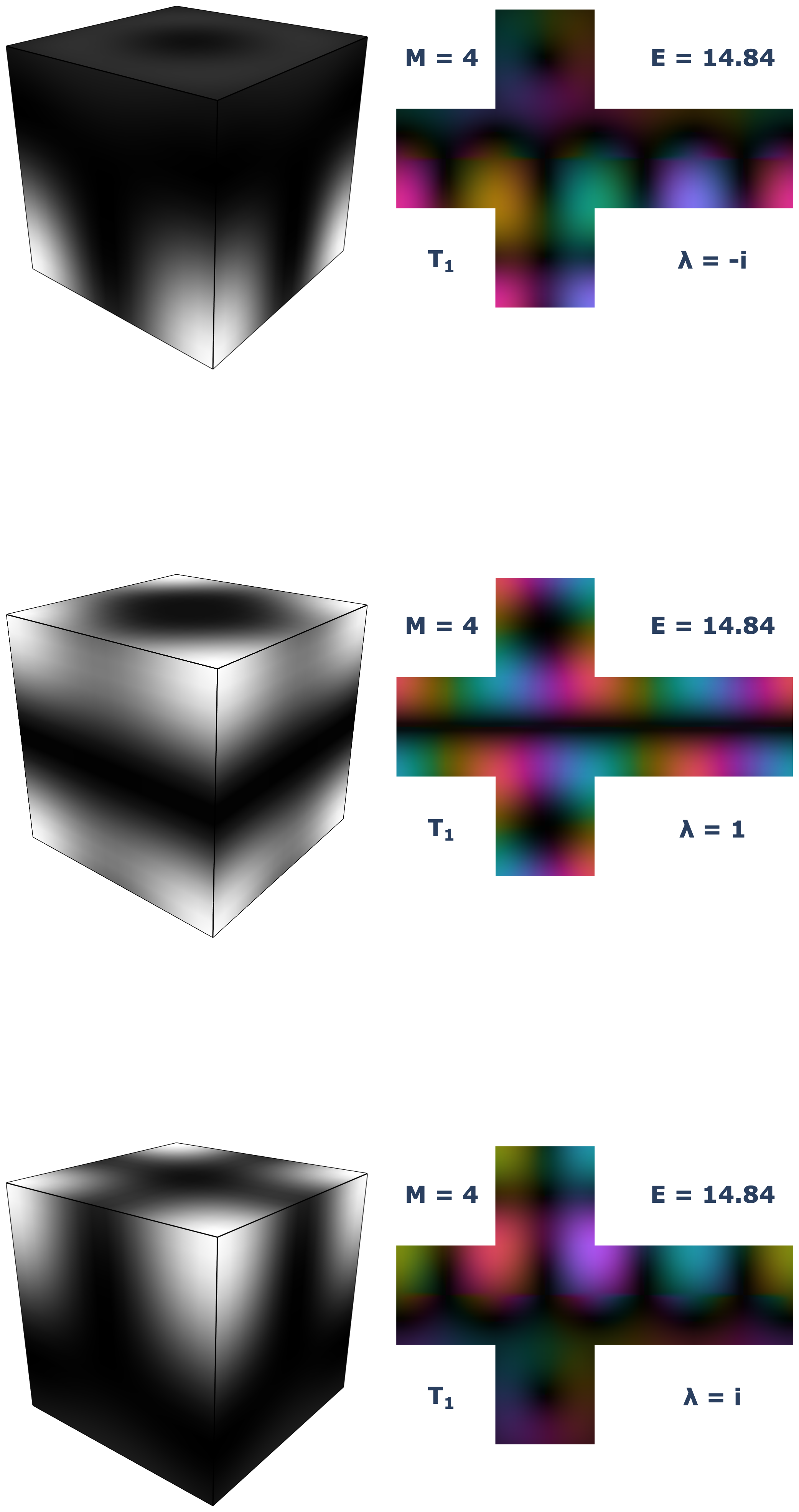}
    \end{minipage}
    \caption{Lowest $T_1$ manifold for $M=0$ (left) and $M=4$ (right). The rows display the complete $C_4$  eigenbasis of the multiplet; the corresponding $\lambda$ values are printed in the panels.}
    \label{fig: M0_M4_T1}
\end{figure*}
\begin{figure*}[hbt!]
    \centering
    \begin{minipage}[t]{0.49\textwidth}
        \centering
        \includegraphics[width=\linewidth]{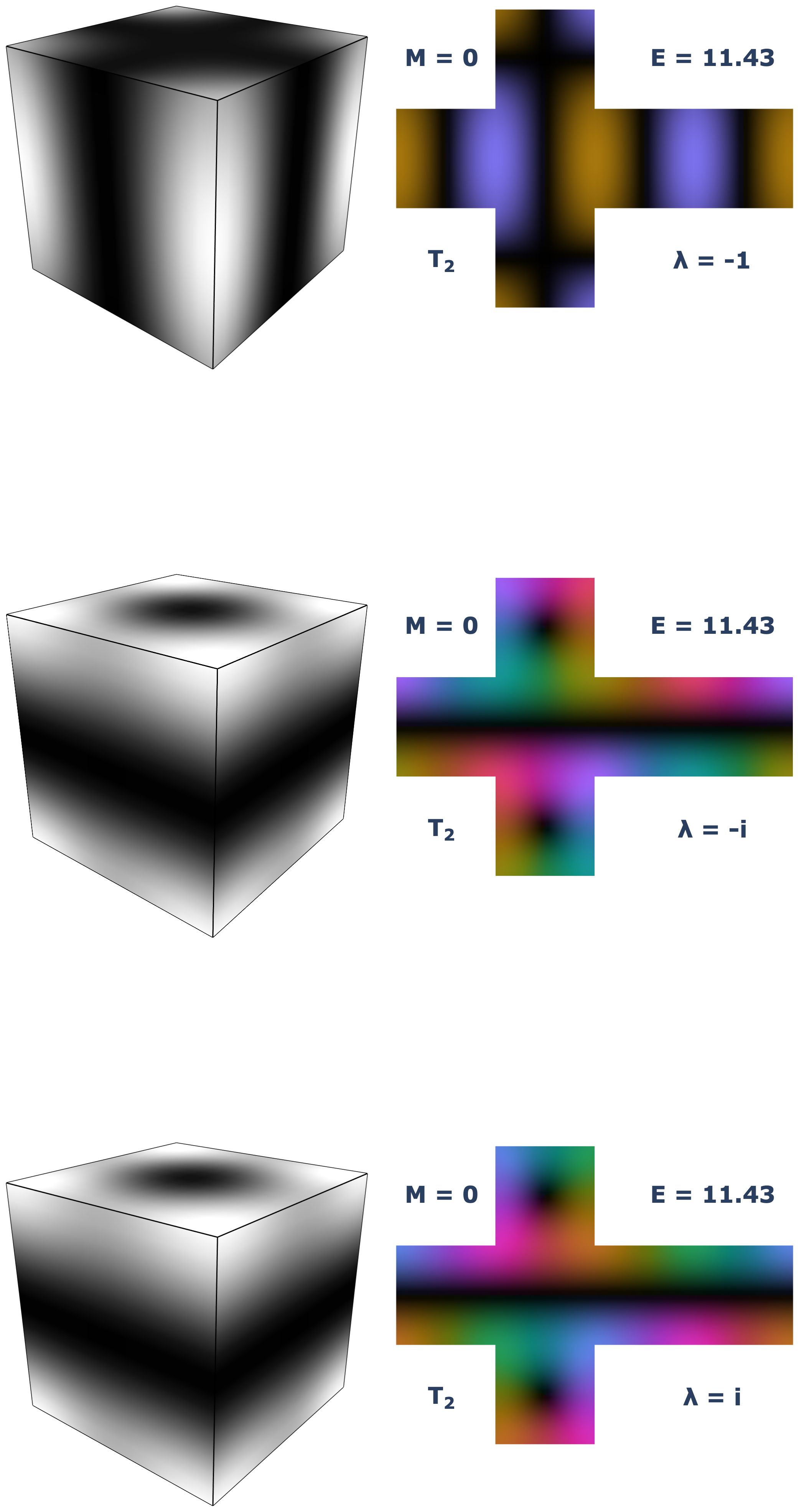}
    \end{minipage}
    \hfill
    \begin{minipage}[t]{0.49\textwidth}
        \centering
        \includegraphics[width=\linewidth]{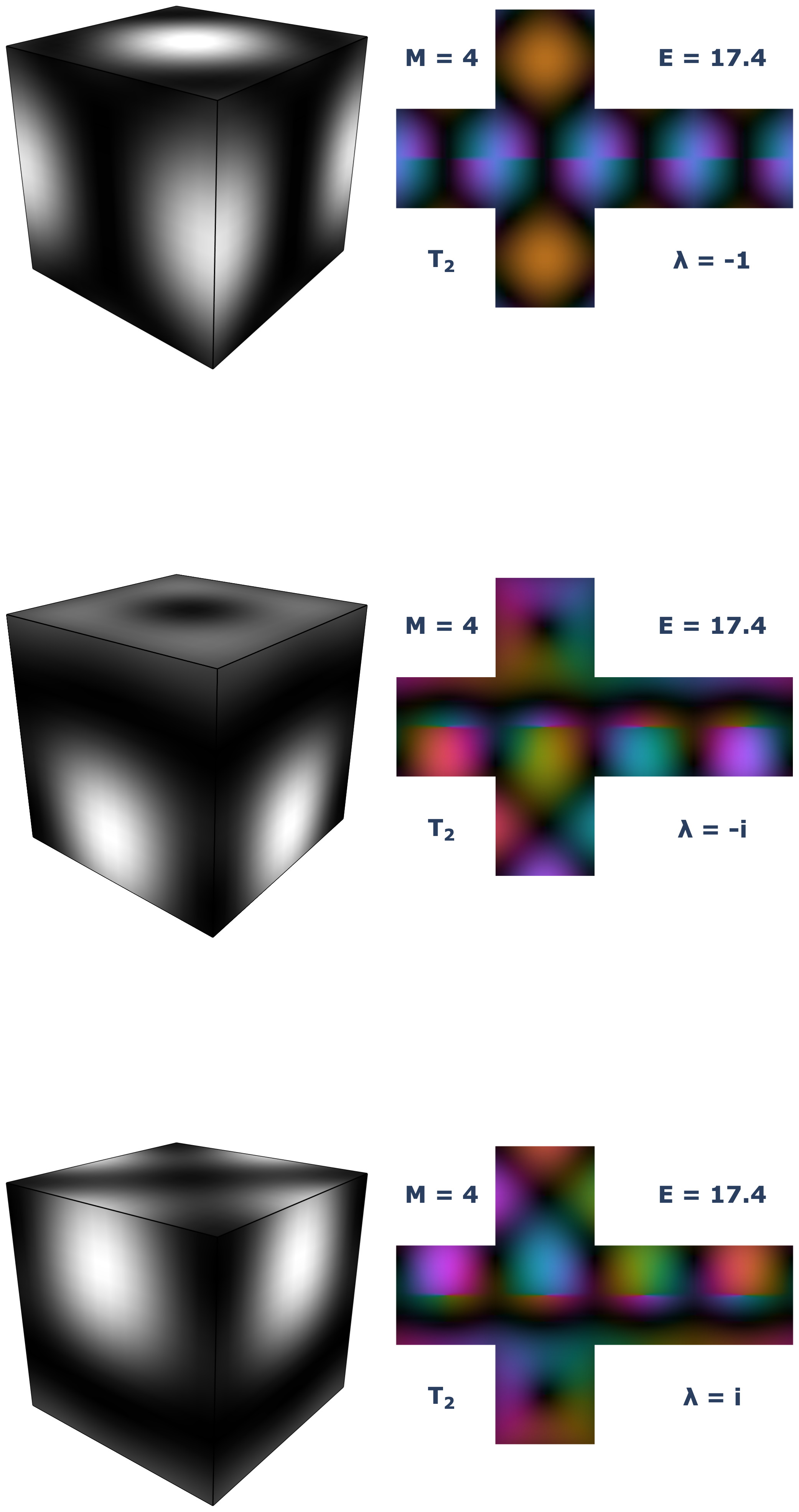}
    \end{minipage}
    \caption{Lowest $T_2$ manifold for $M=0$ (left) and $M=4$ (right). The rows display the complete $C_4$  eigenbasis of the multiplet; the corresponding $\lambda$ values are printed in the panels.}
    \label{fig: M0_M4_T2}
\end{figure*}

\begin{figure*}[hbt!]
    \centering
    \begin{minipage}[t]{0.49\textwidth}
        \centering
        \includegraphics[width=\linewidth]{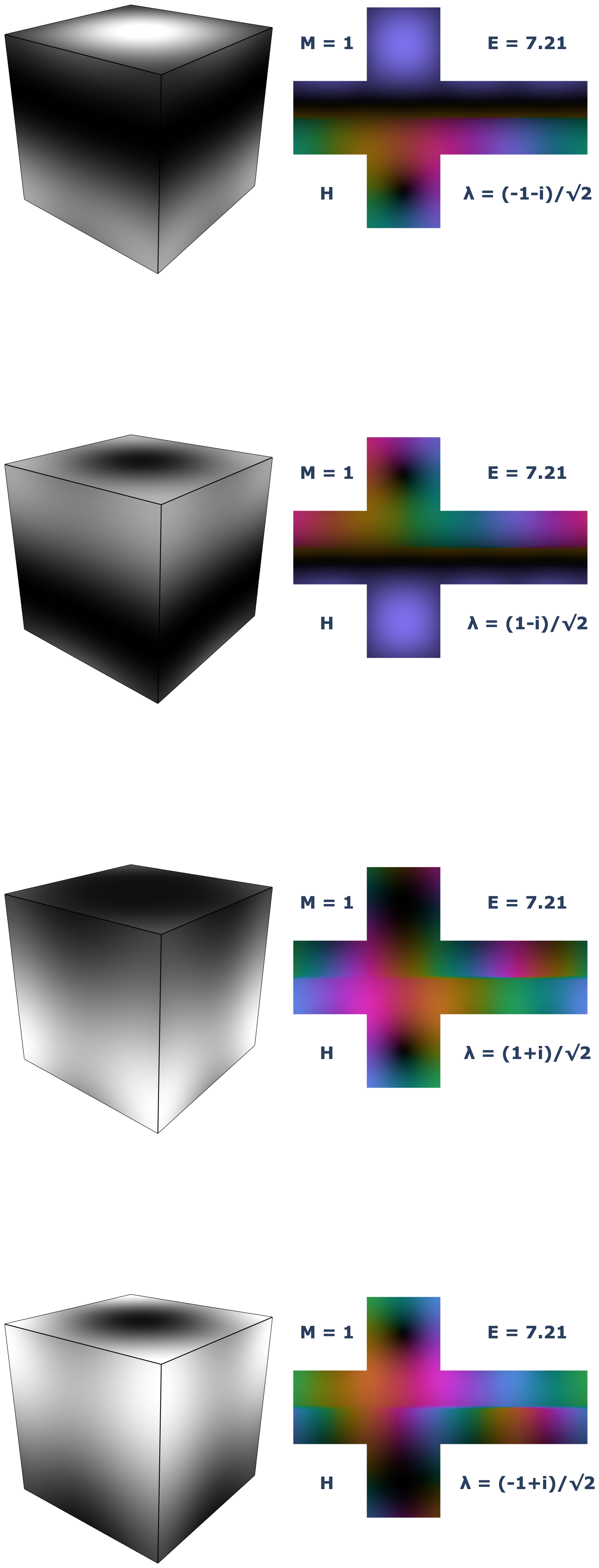}
    \end{minipage}
    \hfill
    \begin{minipage}[t]{0.49\textwidth}
        \centering
        \includegraphics[width=\linewidth]{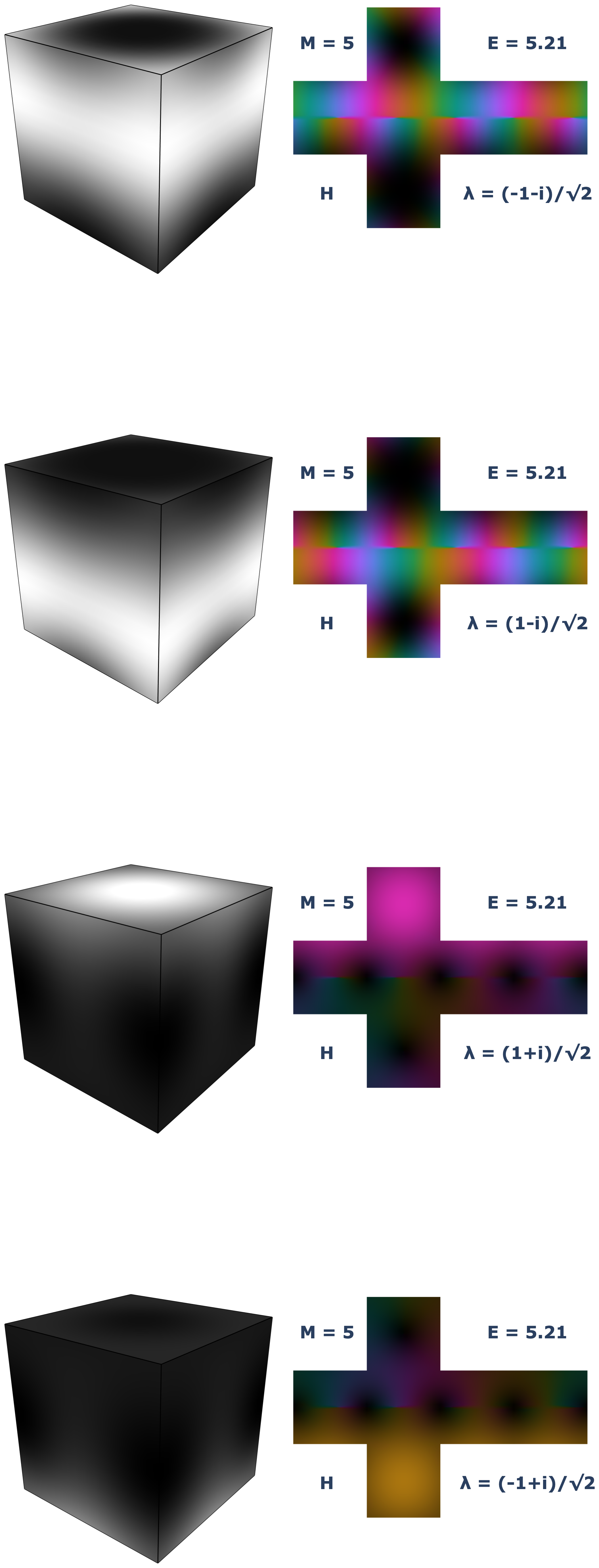}
    \end{minipage}
    \caption{Lowest $H$ manifold for $M=1$ (left) and $M=5$ (right). The rows display the complete $C_4$  eigenbasis of the multiplet; the corresponding $\lambda$  values are printed in the panels.}
    \label{fig: M1_M5_H}
\end{figure*}
\begin{figure*}[hbt!]
    \centering
    \begin{minipage}[t]{0.49\textwidth}
        \centering
        \includegraphics[width=\linewidth]{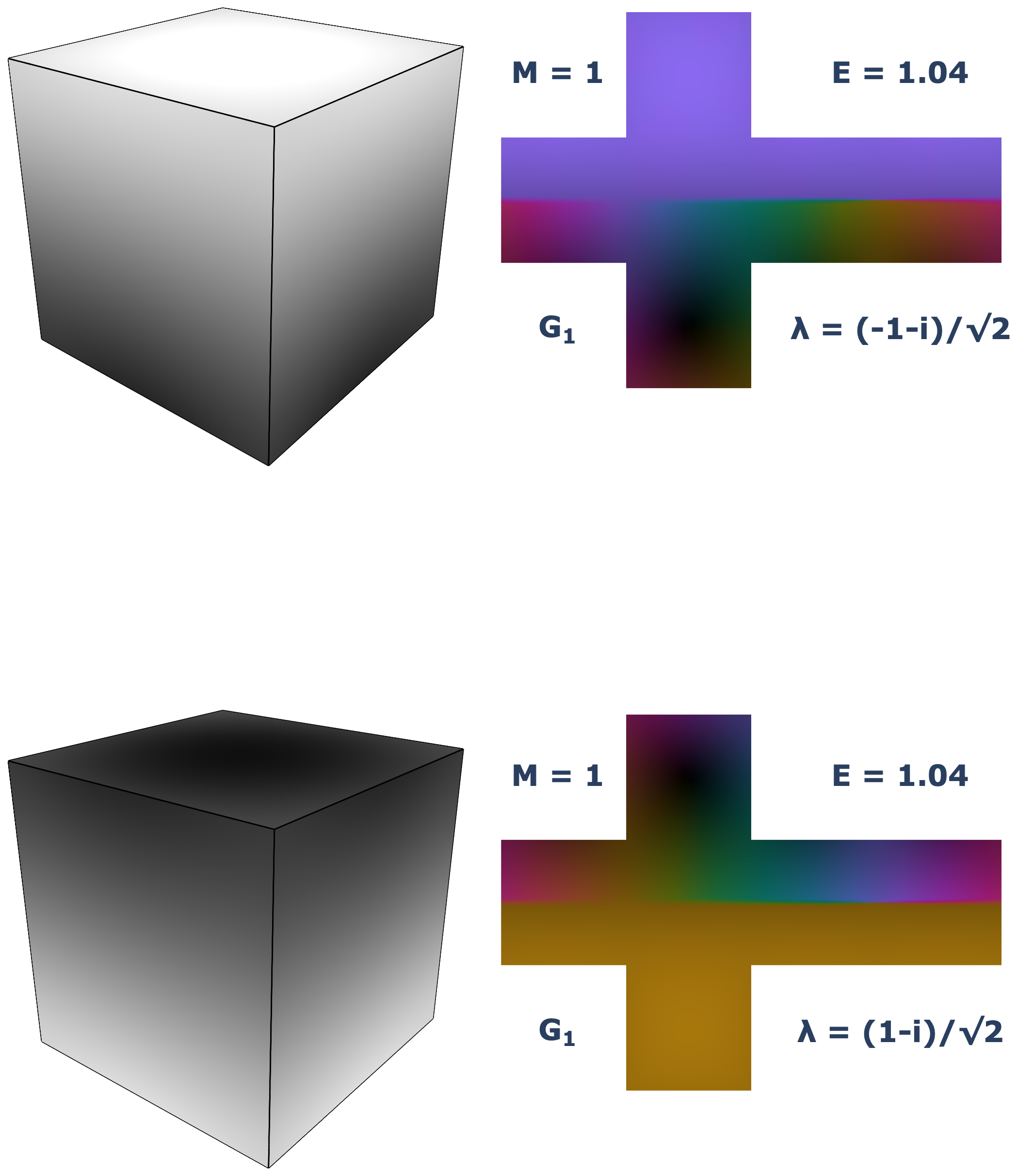}
    \end{minipage}
    \hfill
    \begin{minipage}[t]{0.49\textwidth}
        \centering
        \includegraphics[width=\linewidth]{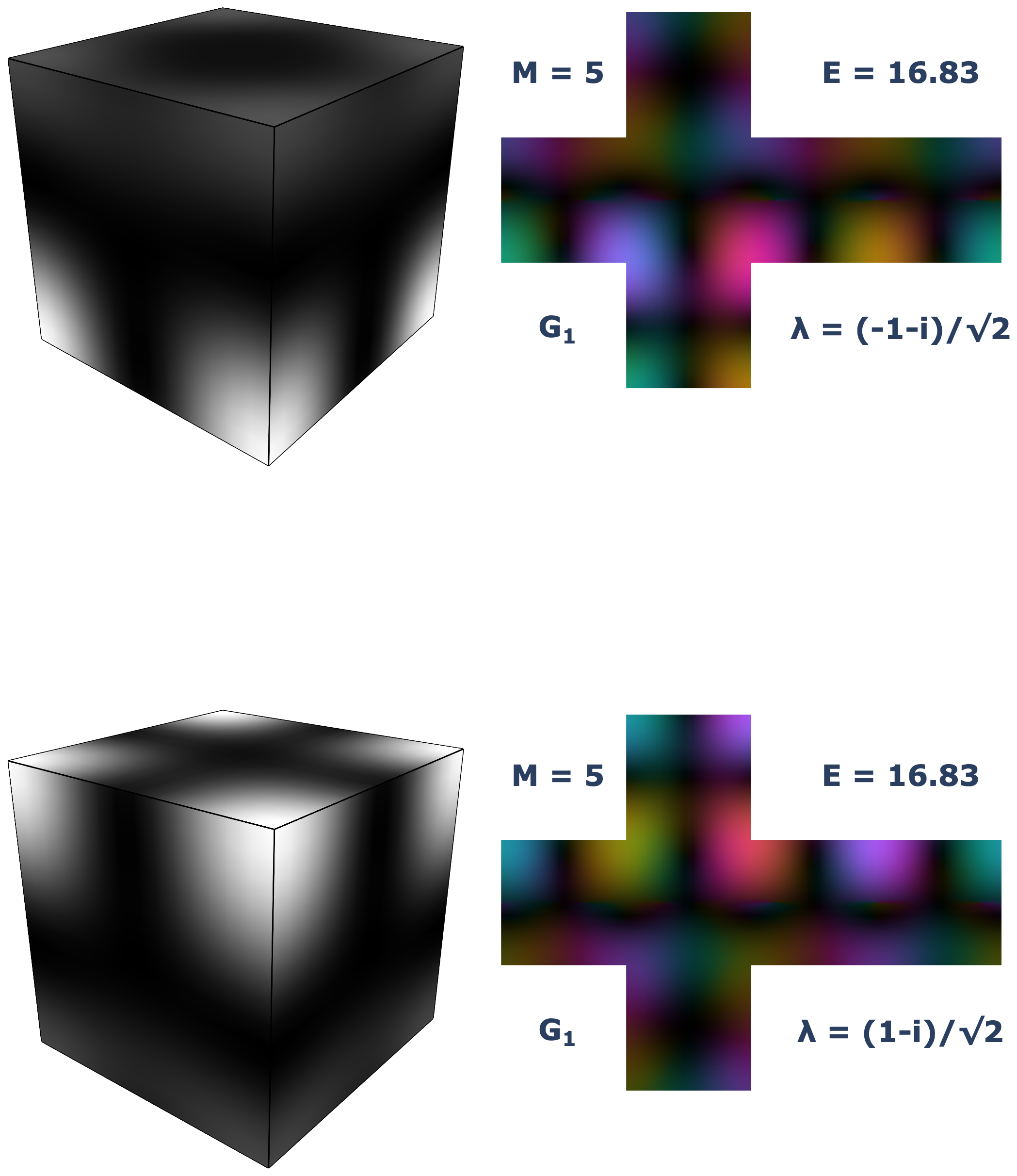}
    \end{minipage}
    \caption{Lowest $G_1$ manifold  for $M=1$ (left) and $M=5$ (right). The rows display the complete $C_4$  eigenbasis of the multiplet; the corresponding $\lambda$ values are printed in the panels.}
    \label{fig: M1_M5_G1}
\end{figure*}
\begin{figure*}[hbt!]
    \centering
    \begin{minipage}[t]{0.49\textwidth}
        \centering
        \includegraphics[width=\linewidth]{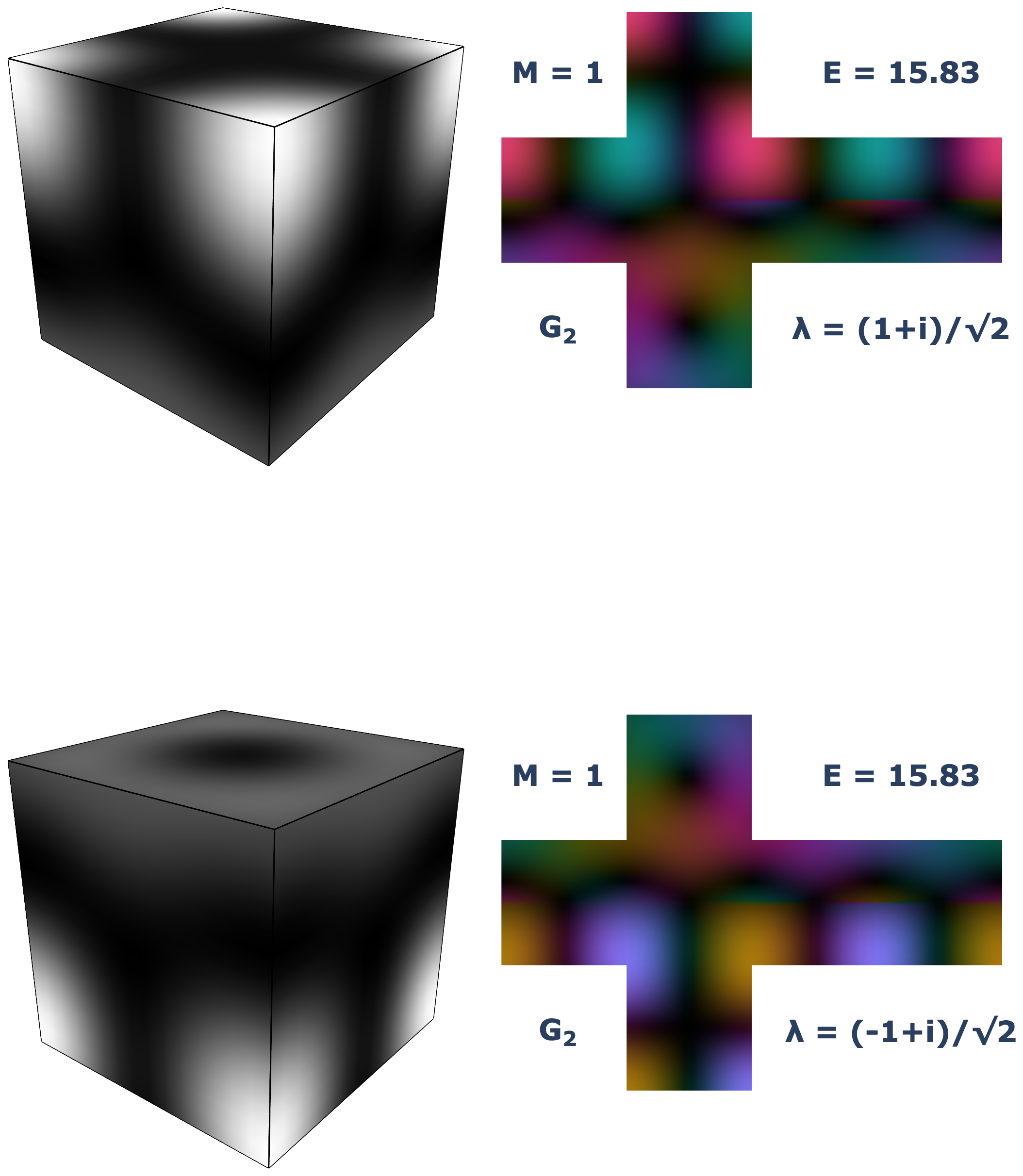}
    \end{minipage}
    \hfill
    \begin{minipage}[t]{0.49\textwidth}
        \centering
        \includegraphics[width=\linewidth]{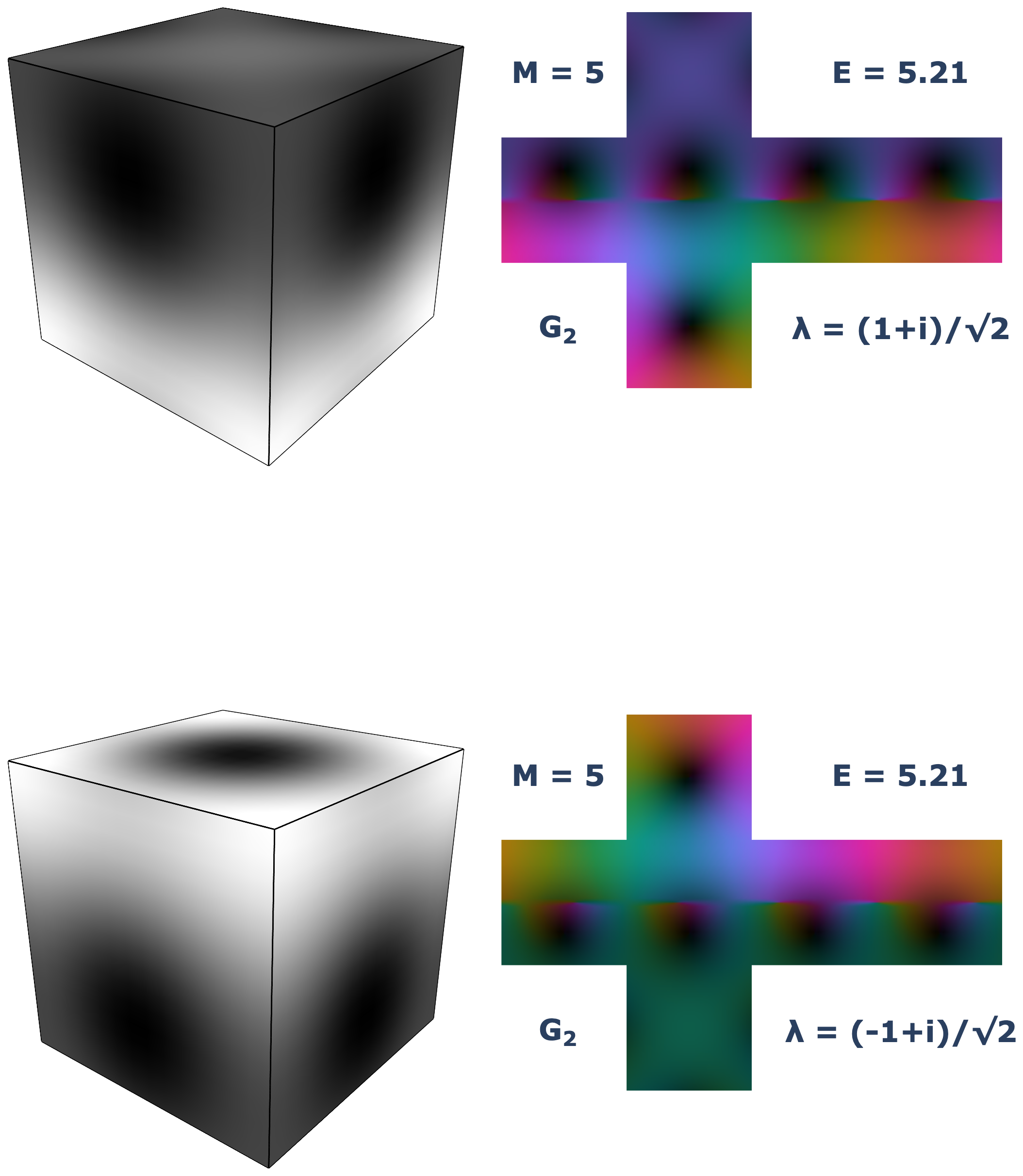}
    \end{minipage}
    \caption{Lowest $G_2$ manifold for $M=1$ (left) and $M=5$ (right). The rows display the complete $C_4$  eigenbasis of the multiplet; the corresponding $\lambda$ values are printed in the panels.}
    \label{fig: M1_M5_G2}
\end{figure*}
\begin{figure*}[hbt!]
    \centering
    \includegraphics[width=1\linewidth]{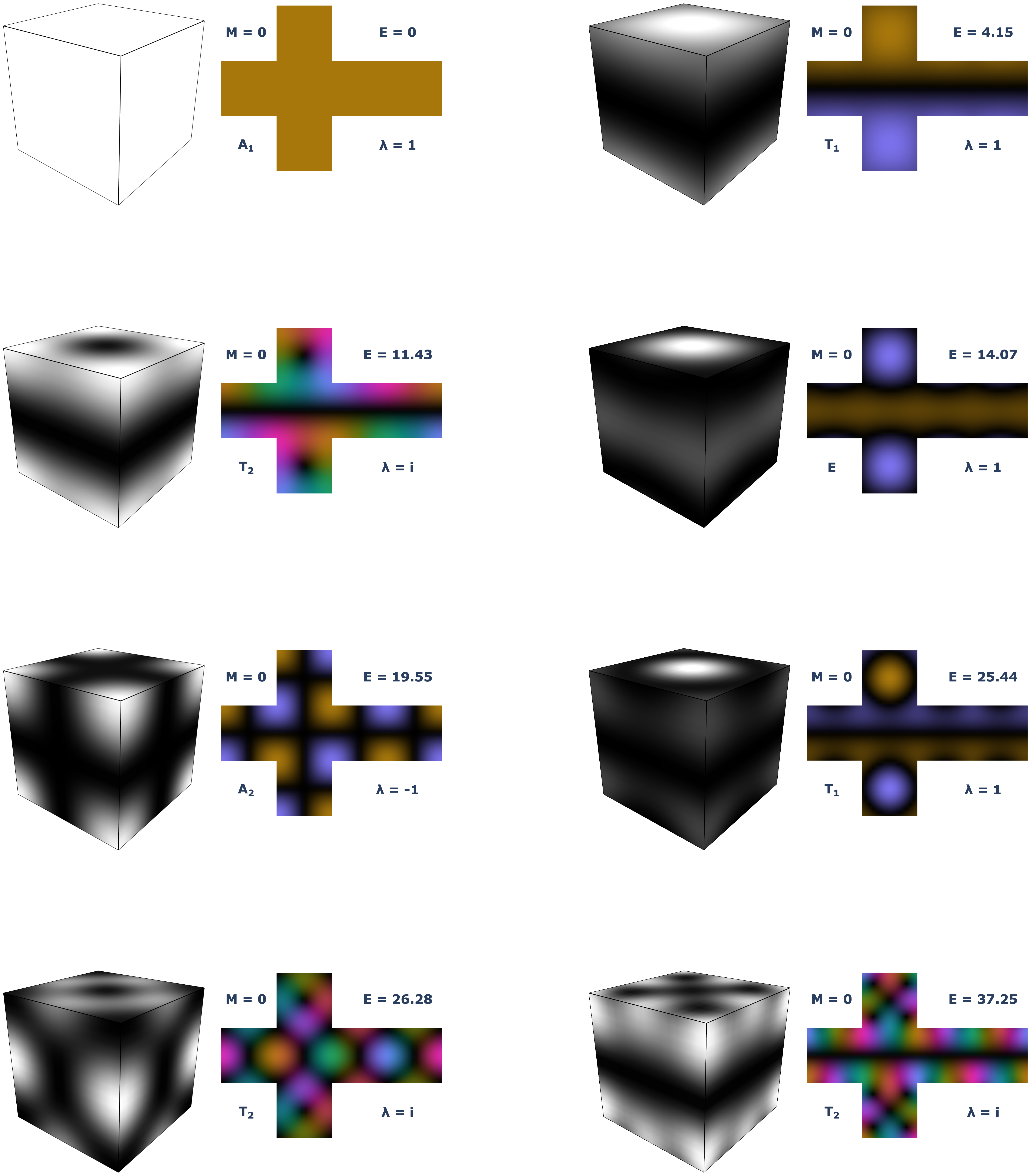}
    \caption{Representative $C_4$ eigenstates from the eight lowest energy multiplets for $M=0$, ordered by increasing energy from upper left to lower right. Each panel gives the energy, irreducible representation, and $C_4$ eigenvalue.  Only one partner wavefunction is displayed for degenerate manifolds.}
    \label{fig: M0_lowest_energies}
\end{figure*}
\begin{figure*}[hbt!]
    \centering
    \includegraphics[width=1\linewidth]{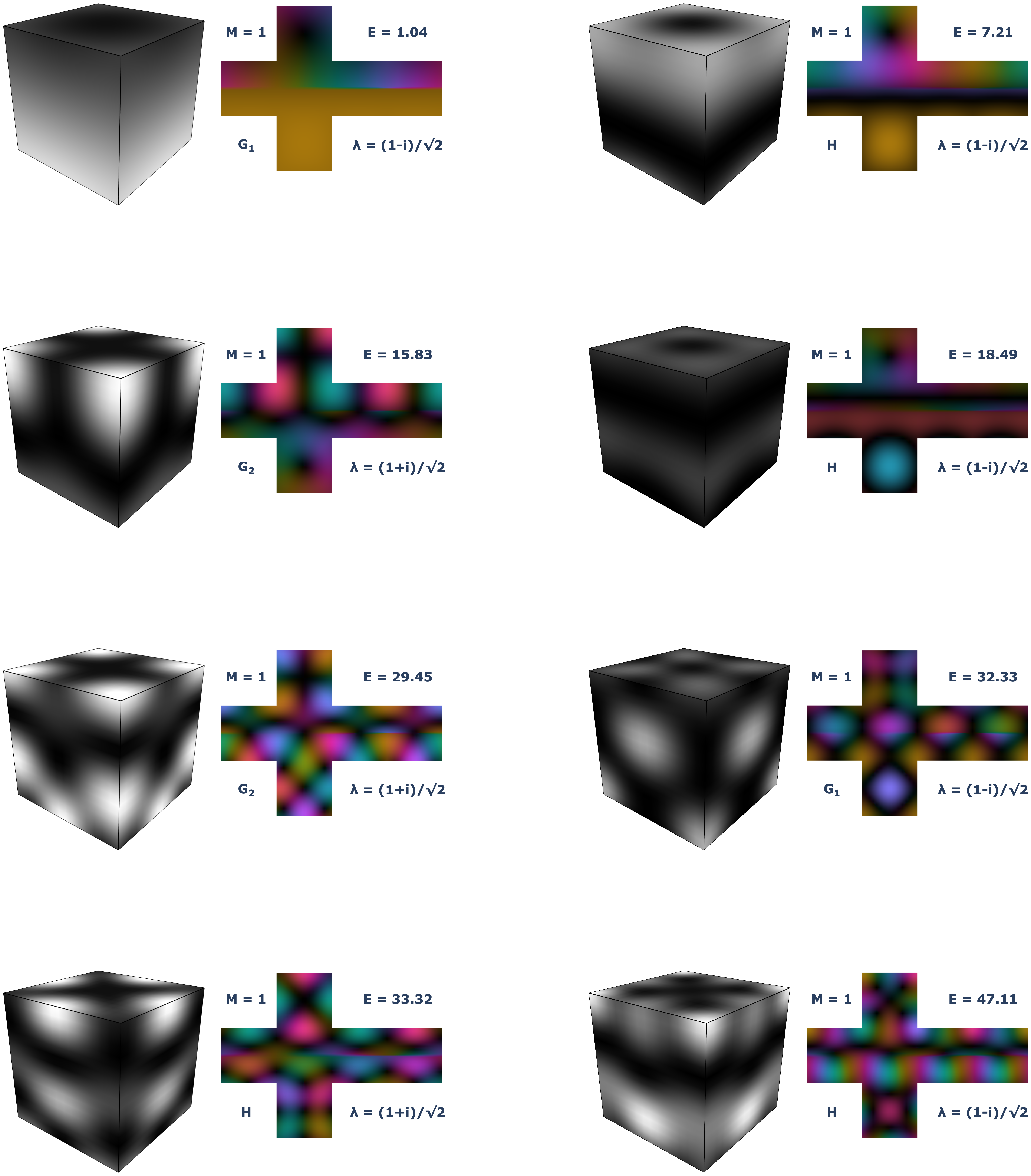}
    \caption{Representative $C_4$ eigenstates from the eight lowest energy multiplets for $M=1$, ordered by increasing energy from upper left to lower right. Each panel gives the energy, irreducible representation, and $C_4$ eigenvalue.  Only one partner wavefunction is displayed for degenerate manifolds.}
    \label{fig: M1_lowest_energies}
\end{figure*}
\begin{figure*}[hbt!]
    \centering
    \includegraphics[width=1\linewidth]{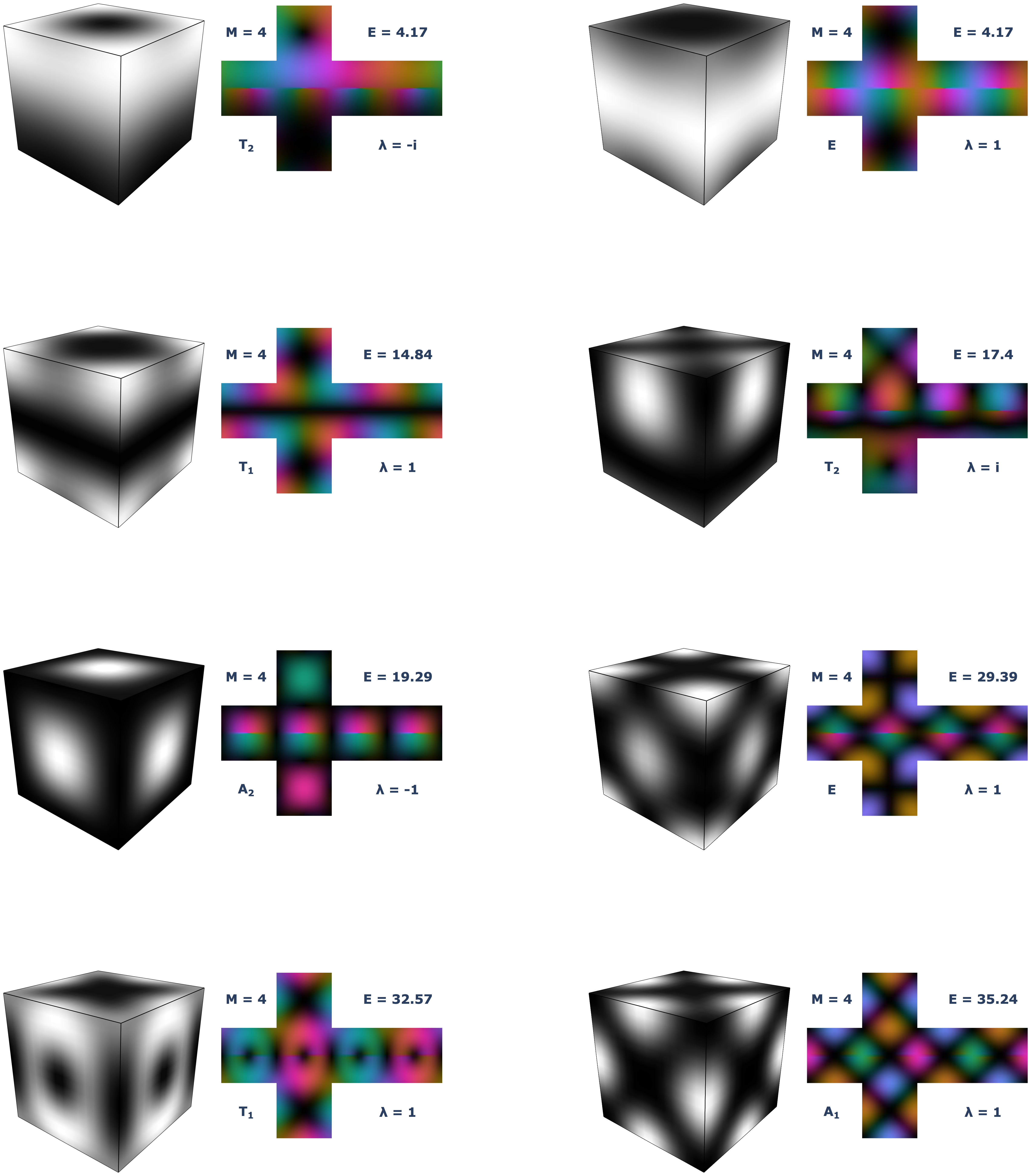}
    \caption{Representative $C_4$ eigenstates from the eight lowest energy multiplets for $M=4$, ordered by increasing energy from upper left to lower right. Each panel gives the energy, irreducible representation, and $C_4$ eigenvalue.  Only one partner wavefunction is displayed for degenerate manifolds.}
    \label{fig: M4_lowest_energies}
\end{figure*}
\begin{figure*}[hbt!]
    \centering
    \includegraphics[width=1\linewidth]{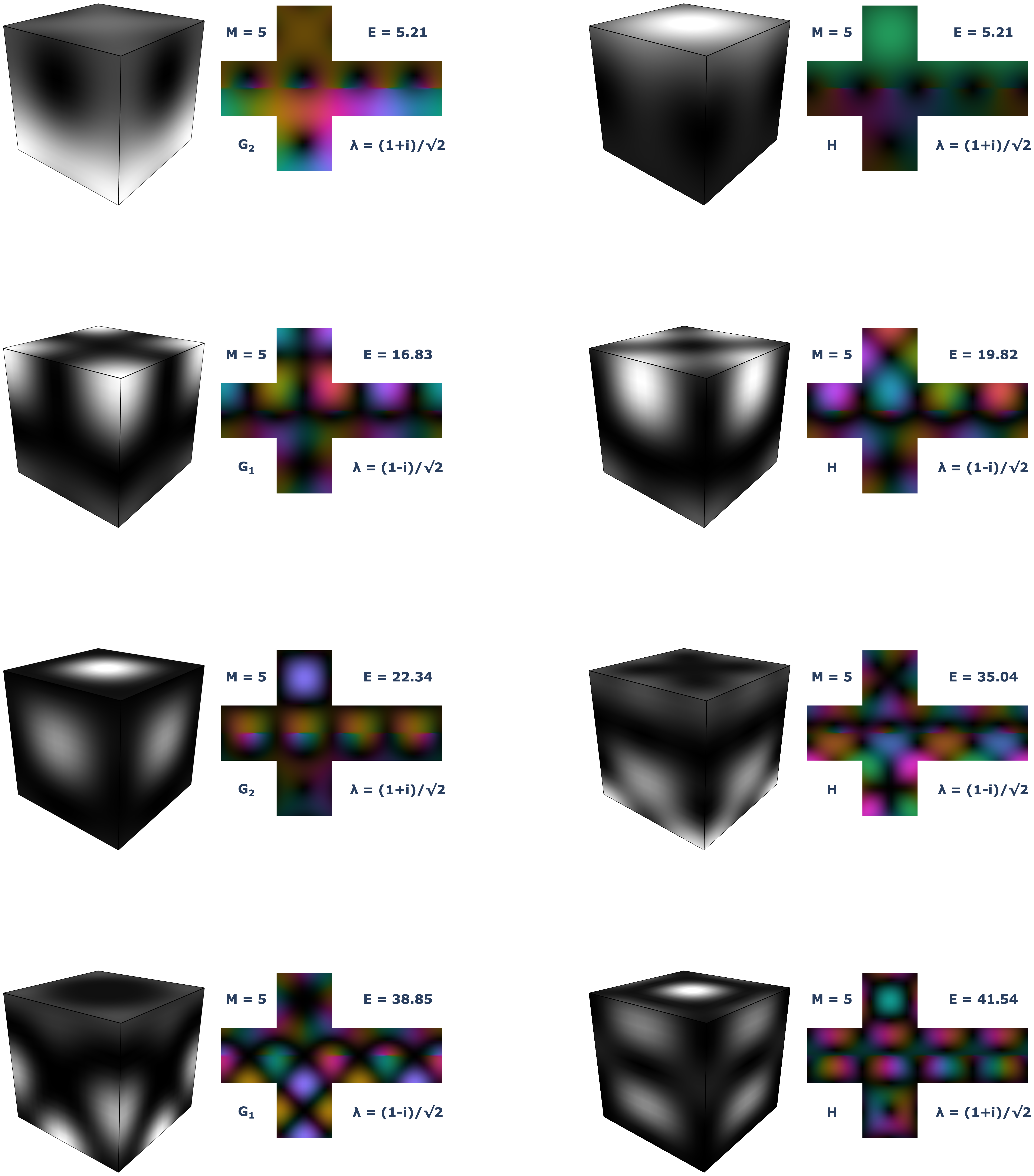}
    \caption{Representative $C_4$ eigenstates from the eight lowest energy multiplets for $M=5$, ordered by increasing energy from upper left to lower right. Each panel gives the energy, irreducible representation, and $C_4$ eigenvalue.  Only one partner wavefunction is displayed for degenerate manifolds.}
    \label{fig: M5_lowest_energies}
\end{figure*}

\bibliography{CubeMonopole.bib}%

\end{document}